# THERMODYNAMICS OF CHLOROBENZENE, OR BROMOBENZENE, OR 1-CHLORONAPHTHALENE OR 1,2,4-TRICHLOROBENZENE + ALKANE MIXTURES

Juan Antonio González,* Luis Felipe Sanz, Fernando Hevia, Isaías. García de la Fuente, and José Carlos Cobos

G.E.T.E.F., Departamento de Física Aplicada, Facultad de Ciencias, Universidad de Valladolid, Paseo de Belén, 7, 47011 Valladolid, Spain.

*corresponding author, e-mail: jagl@termo.uva.es; Fax: +34-983-423136; Tel: +34-983-423757

The systems C$_6$H$_5$Cl, or C$_6$H$_5$Br, or 1-chloronaphthalene, or 1,2,4-trichlorobenzene, or 1-methylnaphthalene, or 1,2,4-trimethylbenzene + alkane have been investigated by means of the their excess molar properties, including, when the needed data are available, those at constant volume, internal energies ($U_{Vm}^E$) and heat capacities ($C_{Vm}^E$), and using the DISQUAC, and Flory models, and the concentration-concentration structure factor formalism. The position of the mixtures within the $G_m^E$ (excess molar Gibbs energy) vs. $H_m^E$ (excess molar enthalpy) diagram has been also determined. Interactions between C$_6$H$_5$X molecules become stronger in the sequence X = H ≈ F ≈ Cl < Br. These interactions are weaker than those between 1-chloronaphtahlene or 1,2,4-trichlorobenzene molecules. It is shown that the considered systems have some common features: dispersive interactions are dominant, structural effects for solutions with shorter *n*-alkanes are large and $U_{Vm}^E$ decreases when the number (*n*) of C atoms of the alkane increases. This variation is held when an *n*-alkane is replaced by a branched alkane with the same *n* in systems with C$_6$H$_5$Cl or 1-chloronaphthalene. This suggests that larger alkanes are poorer breakers of the interactions between aromatic halogenated compounds. Viscosity and $C_{Vm}^E$ data support this conclusión. The parabolic dependence of $C_{Vm}^E$ with *n* indicates that the short orientational order of long *n*-alkanes is destroyed. Aromacity and proximity effects are discussed.

## 1. Introduction

For binary mixtures formed by a molecule A of more or less globular shape and a long *n*-alkane, group contribution models predict molar excess enthalpies, $H_m^E$, at equimolar composition and 298.15 K, which are lower than the experimental results [1,2]. This occurs, e.g. for benzene, or toluene, or cyclohexane, or CCl$_4$ + *n*-alkane systems [1,3,4]. In such cases, the observed differences between theoretical and experimental results have been ascribed to the existence in those solutions of the so-called Patterson's effect. This effect consists in an extra endothermic contribution to the molar excess enthalpies due to the disruption of the existing local order in long *n*-alkanes by A [5-8], in such way that, in order to take into account the mentioned effect, the application of group contribution models requires of interaction parameters dependent on the *n*-alkane [1,3,4]. In a recent work [2], we studied the mixtures C$_6$H$_5$F, or 1,4-C$_6$H$_4$F$_2$ or C$_6$F$_6$ + hydrocarbon and we demonstrated that no Patterson's effect exists in *n*-alkane systems since DISQUAC [9] describes accurately both their $H_m^E(x_1 = 0.5)$ values and the concentration dependence of this excess function using interaction parameters independent of the *n*-alkane. Similarly, UNIFAC (Dortmund) [10] provided good results for mixtures with C$_6$H$_5$F, or 1,4-C$_6$H$_4$F$_2$, although somewhat larger differences between experimental and calculated values were obtained for C$_6$F$_6$ systems [2]. Another relevant result from this study was that DISQUAC correctly reproduces the negative excess molar heat capacities at constant pressure, $C_{pm}^E$. Systems containing *n*-alkane and an aromatic halogenated compound (AHC) such as C$_6$H$_5$X (X = Cl, Br) or 1-chloronaphthalene or 1,2,4-trichlorobenzene form an interesting class of mixtures due to their dependences of $H_m^E(x_1 = 0.5)$ with *n*, the number of C atoms in the *n*-alkane. At 298.15 K, systems with C$_6$H$_5$Cl [11] or C$_6$H$_5$Br [12] show $H_m^E(x_1 = 0.5)$ values which increase smoothly with *n*, while solutions with 1,2,4-trichlorobenzene [13] or 1-chloronaphthalene [14] show decreasing results of this quantity when *n* is increased. Such dependences of $H_m^E(x_1 = 0.5)$ with *n* are rather unusual and have been attributed to the Wilhelm's effect, a typical feature of solutions containing a flat molecule and a long *n*-alkane [15]. It has been pointed out that the Wilhelm's effect might involve an exothermic contribution to $H_m^E$ arising from the fact that the flat molecules hinder the rotational motion of the segments of the flexible molecules of long *n*-alkanes [15,16]. On the other hand, some previous calculations suggest that the application of group contribution models to mixtures where this effect is present requires of interaction parameters which decrease for increased *n* values in order to describe properly the variation of the $H_m^E$ values at equimolar composition and 298.15 K with *n* [12,17]. At this regards, the application of UNIFAC (Dortmund) to C$_6$H$_5$Cl or 1-chloronaphthalene + *n*-alkane mixtures is interesting. In the framework of UNIFAC, both $C_6H_5Cl$ or 1-chloronaphthalene molecules are built by the two main groups ACH (Nº 3) and ACCl (Nº 25) [10]. Thus, mixtures with one of the mentioned molecules and a given solvent (e.g., *n*-alkane) are characterized by the same interaction parameters. In the following, UNIFAC calculations have been conducted using values of the parameters taken from the literature [10]. For $C_6H_5Cl$ mixtures at 298.15 K and $x_1 = 0.5$, we have: $H_m^E$(UNIFAC)/J mol$^{-1}$ = 566 ($n$ = 6), 653 ($n$ = 9) and 740 ($n$ = 14). The experimental results are: (668 ($n$ = 6), 712 ($n$ = 9) and 755 ($n$ = 14)) J mol$^{-1}$ [11]. In the case of systems with 1-chloronaphthalene and at the same conditions, $H_m^E$ (UNIFAC)/J mol$^{-1}$ = 782 ($n$ = 6), 975 ($n$ = 10) and 1149 ($n$ =16) being the experimental values: (665 ($n$ = 6), 654 ($n$ = 10) and 580 ($n$ =16)) J mol$^{-1}$ [14]. We note that the theoretical results are lower than the experimental ones for solutions with $C_6H_5Cl$, while the former are much larger than the latter for mixtures involving 1-chloronaphthalene. Thus, from UNIFAC calculations, no conclusion can be stated about the possible existence of Patterson's or Wilhelm's effects in the considered systems. It seems clear that, at least and as a starting point, different interaction parameters should be used for $C_6H_5Cl$ or 1-chloronaphthalene if one wants to describe properly so subtle effects. For this reason, we investigate the $C_6H_5Cl$ or $C_6H_5Br$ or 1,2,4-trichlorobenzene or 1-chloronaphthalene + *n*-alkane mixtures in terms of DISQUAC since, within this model, it is admitted that the interaction parameters can depend of the molecular structure of the mixture components. As an intermediate and needed step in this research, we also treat similar systems with 1-methylnaphthalene, or 1,2,4-trimethylbenzene. In order to gain a deeper understanding of the main features of this class of systems, excess molar properties at constant volume, internal energy ($U_{Vm}^E$) and heat capacity ($C_{Vm}^E$), are also discussed. The creation/destruction of order in systems with chlorobenzene [18] or 1-chloronaphthalene [16] + alkanes has been investigated in detail comparing experimental second-order thermodynamic mixing properties at equimolar composition with theoretical results from the Flory model [19]. However, in those studies the mentioned excess molar properties at constant volume were not considered, perhaps due to, in the framework of the theory, $C_{Vm}^E = 0$. Here, we also apply the Flory model to investigate orientational effects in the selected systems. In addition, we examine important effects such proximity effects between the phenyl ring and the Cl atom in systems involving molecules of the type $C_6H_5$-$(CH_2)_u$-Cl and the aromacity effect. Finally, the systems are studied by means of the concentration-concentration structure factor [20-22], a method concerned with the study of fluctuations in the number of molecules in a given binary mixture regardless of the components, the fluctuations in the mole fraction and the cross fluctuations.

Halogenated compounds have a large number of applications. In Medical Chemistry, this is due to the halogen bonding that they can form [23-25]. Chloronaphthalenes have applications in the industry as heat transfer media, lubricant additives or electric insulating materials.

## 2. THEORIES

### 2.1 DISQUAC

DISQUAC is based on the Guggenheim´s rigid lattice theory [26]. We highlight some relevant features of the model. (i) The geometrical parameters, the total molecular volumes, $r_i$, surfaces, $q_i$, and the molecular surface fractions, $\alpha_i$, of the mixture compounds are calculated additively on the basis of the group volumes $R_G$ and surfaces $Q_G$ recommended by Bondi [27]. As volume and surface units, the volume $R_{CH4}$ and surface $Q_{CH4}$ of methane are taken arbitrarily [28]. The geometrical parameters for the groups considered along the study are available in the literature [28-31]. (ii) The partition function is factorized into two terms. Thus, the excess functions are the result of the sum of a dispersive (DIS) term linked to dispersive forces, and of a quasichemical (QUAC) term arising from the anisotropy of the field forces created by the solution molecules. In the case of the molar excess Gibbs energy, $G_m^E$, a combinatorial term, $G_m^{E,COMB}$, represented by the Flory-Huggins equation [28] must be included:

$$G_m^E = G_m^{E,COMB} + G_m^{E,DIS} + G_m^{E,QUAC} \quad (1)$$

$$H_m^E = H_m^{E,DIS} + H_m^{E,QUAC} \quad (2)$$

(iii) The interaction parameters are considered that depend on the molecular structure. (iv) The same coordination number $z = 4$ is used for all the polar contacts. This is a key shortcoming of the model and is partially removed using interaction parameters dependent on the molecular structure. (v) It is also assumed that $V_m^E$ (excess molar volume) = 0. Therefore, no volume change is produced upon mixing.

The equations required to calculate the DIS and QUAC contributions to $G_m^E$ and $H_m^E$ can be found elsewhere [9,32]. The temperature dependence of the interaction parameters is expressed in terms of the DIS and QUAC interchange coefficients [9,32], $C_{st,l}^{DIS}; C_{st,l}^{QUAC}$ where s ≠ t are two contact surfaces present in the mixture and $l$ = 1 (Gibbs energy; $C_{st,1}^{DIS/QUAC} = g_{st}^{DIS/QUAC}(T_o)/RT_o$ ); $l$ = 2 (enthalpy, $C_{st,2}^{DIS/QUAC} = h_{st}^{DIS/QUAC}(T_o)/RT_o$ )); $l$ = 3 (heat capacity, $C_{st,3}^{DIS/QUAC} = c_{pst}^{DIS/QUAC}(T_o)/R$)). $T_o$ = 298.15 K is the scaling temperature and $R$, the gas constant.

*2.2    Flory model*

A short summary of the main hypotheses of the theory [19, 33-36] can be encountered elsewhere [37] and will not be repeated here. We merely remark that random mixing is a basic assumption of the model.   The Flory equation of state is:

$$\frac{\bar{P}\bar{V}}{\bar{T}} = \frac{\bar{V}^{1/3}}{\bar{V}^{1/3}-1} - \frac{1}{\bar{V}\bar{T}} \quad (3)$$

where $\bar{V} = V_m/V_m^*$; $\bar{P} = P/P^*$ and $\bar{T} = T/T^*$ are the reduced properties: volume, pressure and temperature, respectively. Equation (3) is valid for pure liquids and liquid mixtures. In the former case, the reduction parameters, $V_{mi}^*$, $P_i^*$ and $T_i^*$ are determined from densities, $\rho_i$, isobaric thermal expansion coefficients, $\alpha_{pi}$, and isothermal compressibilities, $\kappa_{Ti}$, data. Expressions for reduction parameters of mixtures are given elsewhere [37]. $H_m^E$ values are determined from:

$$H_m^E = \frac{x_1 V_{m1}^* \theta_2 X_{12}}{\bar{V}} + x_1 V_{m1}^* P_1^* (\frac{1}{\bar{V}_1} - \frac{1}{\bar{V}}) + x_2 V_{m2}^* P_2^* (\frac{1}{\bar{V}_2} - \frac{1}{\bar{V}}) \quad (4)$$

where all the symbols have their usual meaning [37]. The reduced volume of the mixture, $\bar{V}$, in equation (4) is obtained from (3). That is, $V_m^E$ can be also calculated:

$$V_m^E = (x_1 V_{m1}^* + x_2 V_{m2}^*)(\bar{V} - \varphi_1 \bar{V}_1 - \varphi_2 \bar{V}_2) \quad (5)$$

*2.3 The concentration-concentration structure factor formalism*

For binary systems, the $S_{CC}(0)$ function can be computed from the expression [20,21]:

$$S_{CC}(0) = \frac{x_1 x_2}{1 + \frac{x_1 x_2}{RT}\left(\frac{\partial^2 G_m^E}{\partial x_1^2}\right)_{P,T}} \quad (6)$$

In the case of an ideal system, $G_m^{E,id} = 0$; and $S_{CC}^{id}(0) = x_1 x_2$. Stability conditions require that $S_{CC}(0) > 0$. Thus, if a mixture is close to phase separation, $S_{CC}(0)$ must be large and positive, the dominant trend is then the separation between components (homocoordination), and $S_{CC}(0) > x_1x_2$. If compound formation between components is dominant (heterocoordination), $S_{CC}(0)$ must be very low and $0 < S_{CC}(0) < x_1x_2$. Additional details are available in reference [20]. In terms of DISQUAC and taking into account equation (1), it is possible to write [38]:

$$\frac{1}{S_{CC}(0)} = \frac{1}{x_1x_2} + \frac{1}{RT}\left(\frac{\partial^2 G^{E,COMB}}{\partial x_1^2}\right)_{P,T} + \frac{1}{RT}\left(\frac{\partial^2 G^{E,int}}{\partial x_1^2}\right)_{P,T} \quad (7)$$

$$= \frac{1}{x_1x_2} + S_{CC}(0)^{-1}(\text{comb}) + S_{CC}(0)^{-1}(\text{int})$$

where $G_m^{E,int} = G_m^{E,DIS} + G_m^{E,QUAC}$. This makes possible to evaluate size effects on $S_{CC}(0)$.

### 3. Estimation of interaction parameters

*3.1 DISQUAC*

In the framework of DISQUAC, $C_6H_5X$ (X = Cl, Br) or 1-chloronaphthalene or 1,2,4-trichlorobenzene, or 1-methylnaphthalene, or 1,2,4-trimethylbenzene + *n*-alkane, or + cyclohexane mixtures are built by the following types of contacts: (i) type s, s = a, aliphatic ($CH_3$, $CH_2$, in *n*-alkanes, 1-methylnaphthalene, or 1,2,4-trimethylbenzene); or s = c, (c-$CH_2$ in cyclohexane); (ii) type b, aromatic ($C_6H_5$-, in chlorobenzene or bromobenzene; $C_6H_3$-, in 1,2,4-trichlorobenzene or 1,2,4-trimethylbenzene; $C_{10}H_7$-, in 1–chloronaphthalene or 1-methylnaphthalene); (iii) type g, g = Cl (in aromatic chlorinated compounds), or g = Br in $C_6H_5Br$. The general procedure applied when estimating the DISQUAC interaction parameters have been given elsewhere [9,32]. Some important remarks are provided below.

*3.1.1 The 1-methylnaphthalene or 1,2,4-trimethylbenzene + n-alkane systems*

These mixtures are built only by the (a,b) contact, whose interaction parameters must be determined before proceeding to the estimation of the interaction parameters of mixtures with 1-chloronaphthalene, or 1,2,4-trichlorobenzene. Due to 1-methylnaphthalene and 1,2,4-trimethylbenzene are compounds of low dipole moment ($\mu = 0.30$ and 0.51 D, respectively [39]), we have assumed that the (a,b) contact is represented only by dispersive parameters (Table S1, supplementary material). Benzene, or alkyl-benzenes + *n*-alkane mixtures have been treated similarly [29].

*3.1.2 The aromatic halogenated compound + alkane systems*

Three types of contacts exist in these mixtures: (s,b), (s,g) and (b,g), where s = a, aliphatic in *n*-alkanes; or s = c, c-$CH_2$ in cyclohexane. The (s,b) contacts are described by dispersive interaction parameters only and are known from the study of alkyl-benzene + alkane mixtures for systems with toluene [29]. The parameters corresponding to mixtures with 1-

methylnaphtahlene or 1,2,4-trimethylbenzene have been obtained in this work (see above). The interaction parameters for the (b,g) contacts should be determined on the basis of $G_m^E$ and $H_m^E$ data for AHC + benzene mixtures. In the case of systems with 1-chloronaphthalene or 1,2,4-trichlorobenzene, the ThermoLit application showed that the data required to fit the parameters are not available. Mixtures including $C_6H_5X$ (X = Cl, Br) are characterized by low $G_m^E$ and $H_m^E$ values. Thus, at 298.15 K and equimolar composition, for the $C_6H_5Cl$ system $G_m^E$ = 25 J mol$^{-1}$ [40] and $H_m^E$ = −5.1 J mol$^{-1}$ [41], while the latter excess function is 24 J mol$^{-1}$ for the solution with $C_6H_5Br$ [41]. These experimental results suggest that the interaction parameters are merely dispersive. For the sake of simplicity, we have assumed that $C_{bg,l}^{DIS}$ (l =1,2,3) = 0 for all the systems. Therefore only the parameters for the (a,g) contacts have to be determined (Table 1). These contacts were considered to be represented by DIS and QUAC parameters since the aromatic halogenated compounds under study are polar molecules. Their dipole moments are: 1.58 D ($C_6H_5Cl$); 1.57 D ($C_6H_5Br$); 1.26 D (1,2,4-trichlorobenzene); 1.52 D (1-chloronaphthalene) [39]. The adjustment of the (c,g) parameters was conducted similarly assuming that that $C_{ag,l}^{QUAC} = C_{cg,l}^{QUAC}$ (l =1,2,3). This condition is systematically applied in the DISQUAC characterization of mixtures formed by a polar compound and *n*-alkane or cyclohexane [30, 42-46].

*3.2 Estimation of the Flory interaction parameter*

$X_{12}$ can be determined from a $H_m^E$ measurement at given composition from [47,48]:

$$X_{12} = \frac{x_1 P_1^* V_{m1}^* (1 - \frac{\overline{T_1}}{\overline{T}}) + x_2 P_2^* V_{m2}^* (1 - \frac{\overline{T_2}}{\overline{T}})}{x_1 V_{m1}^* \theta_2} \qquad (8)$$

In order to apply this expression, we note that $\overline{VT}$ is a function of $H_m^E$:

$$H_m^E = \frac{x_1 P_1^* V_{m1}^*}{\overline{V_1}} + \frac{x_2 P_2^* V_{m2}^*}{\overline{V_2}} + \frac{1}{\overline{VT}} (x_1 P_1^* V_{m1}^* \overline{T_1} + x_2 P_2^* V_{m2}^* \overline{T_2}) \qquad (9)$$

and that, from the equation of state, $\overline{V} = \overline{V}(\overline{T})$. For more details, see, e.g., references [47,48]. The molar volumes, $V_{mi}$, $\alpha_{Pi}$, $\kappa_{Ti}$, and the corresponding reduction parameters, $P_i^*$ and $V_i^*$, at 298.15 K of AHC, 1-methylnaphthalene and 1,2,4-trimethylbenzene, needed for calculations, are listed in Table 2. For *n*-alkanes, the values used are the same that in the application of the Flory model to alkanone, alkanal or dialkyl carbonate + alkane systems [49]. $X_{12}$ values determined from experimental $H_m^E$ data at $x_1 = 0.5$ and 298.15 K are collected in Tables 1 and S1.

## 4. Theoretical results

Results from DISQUAC on phase equilibria, $H_m^E$ and $C_{pm}^E$ are shown in Tables 3-6 and in Figures 1-3 and S1-S3 for mixtures with AHC, and in Tables S2 and S3 for systems involving 1-methylnaphthalene, or 1,2,4-trimethylbenzene. Tables 3 and 4 contain relative deviations for pressure and for $H_m^E$, respectively, defined as

$$\sigma_r(P) = \{\frac{1}{N}\sum\left[\frac{P_{exp} - P_{calc}}{P_{exp}}\right]^2\}^{1/2} \tag{10}$$

$$dev(H_m^E) = \{\frac{1}{N}\sum\left[\frac{H_{m,exp}^E - H_{m,calc}^E}{H_{m,exp}^E(x_1 = 0.5)}\right]^2\}^{1/2} \tag{11}$$

where $N$ stands for the number of data points. Similarly, results on $H_m^E$ from the application of the Flory model are listed in Tables 4 and S2. Table S4 compares experimental $V_m^E$ values at equimolar composition with Flory calculations. DISQUAC reproduces well vapour-liquid equilibria at high temperatures for the systems 1-methylnaphthalene + n-decane [50] ($T = 523.15$ K; $\sigma_r(P) = 0.070$; experimental value, 0.023), or 1,2,4-trimethylbenzene + n-decane [51,52] ($T = 383.15$ K; $\sigma_r(P) = 0.005$; experimental result, 0.003). The available experimental data on solid-liquid equilibria for the 1,2,4-trichlorobenzene + n-alkane systems [53] are also well represented by the model (Table 5, Figure S3). Calculations have been conducted according to the well-known equation [54]:

$$-\ln x_1 = (\Delta H_{m1}/R)[1/T - 1/T_{m1}] - (\Delta C_{Pm1}/R)[\ln(T/T_{m1}) + (T_{m1}/T) - 1] + \ln \gamma_1 \tag{12}$$

In this equation $x_1$ is the mole fraction and $\gamma_1$ the activity coefficient of component 1 in the solvent mixture, at temperature $T$. Values of $\gamma_1$ were calculated according to DISQUAC using the values of the physical constants ($T_m$, melting temperature; $\Delta H_m$, enthalpy of fusion, $\Delta C_{pm}$, molar heat capacity change during the melting process of component 1) given in reference [53]. DISQUAC results on $H_m^E$ are slightly better than those obtained from the Flory model. This theory represents the main features of $V_m^E$: the increasing of the function with $n$ in systems with a given AHC or 1,2,4-trimethylbenzene, and the large negative values which are typical of solutions containing shorter $n$-alkanes (Table S4). The dependence of the $C_{pm}^E$ data with $n$ for mixtures with $C_6H_5Cl$, 1-chloronaphthalene or 1,2,4-trichlorobenzene is correctly described by DISQUAC using $C_{ag,3}^{DIS}$ values dependent of the $n$-alkane (Table 6). Table 7 contains DISQUAC calculations on $S_{CC}(0)$, and a short comparison between theoretical results with experimental values determined in this work using VLE data from the literature. We note that DISQUAC is a reliable tool to evaluate the $S_{CC}(0)$ magnitude (see also Figure 4).

### 5. Discussion

In the following, values of the thermodynamic properties are given at equimolar composition and 298.15 K.

Aromatic halogenated compound + alkane systems show positive $H_m^E$ values (Table 4 and Figs 1-3, see below) and, therefore, are mainly characterized by interactions between AHC molecules. It is to be noted that, within the diagram $G_m^E$ vs. $H_m^E$ [55-57], mixtures with $C_6H_5Cl$ or $C_6H_5Br$ are situated close to the line $G_m^E = 1/2\, H_m^E$. That is, they are placed near to the region where systems characterized by dispersive interactions are encountered (the region between the lines $G_m^E = 1/3\, H_m^E$ and $G_m^E = 1/2\, H_m^E$). For heptane solutions, $G_m^E$/J mol$^{-1}$ = 446 ($C_6H_5Cl$); 479 ($C_6H_5Br$) (DISQUAC values, this work) and $H_m^E$/J mol$^{-1}$ = 672 ($C_6H_5Cl$) [11]; 795 ($C_6H_5Br$) [12] and $(G_m^E/H_m^E)$ = 0.66 ($C_6H_5Cl$); 0.60 ($C_6H_5Br$). In addition, the $C_{pm}^E$ values are negative: ($-0.75$ and $-3.52$) J mol$^{-1}$ K$^{-1}$ for the mixtures $C_6H_5Cl$ + heptane [58], or + hexadecane [59], respectively, and the corresponding $TS_m^E (= H_m^E - G_m^E)$ values are positive. Such results allow conclude that dispersive interactions are dominant in the mentioned mixtures. The same occurs in $C_6H_5F$ + $n$-alkane systems, which is supported by their $C_{pm}^E$ values, ($-1.19$ ($n$ = 7); $-3.75$ ($n$ = 14)) J mol$^{-1}$ K$^{-1}$ [60], and by their situation in the diagram $G_m^E$ vs. $H_m^E$ (($G_m^E/H_m^E$) = 0.55 for the solution containing heptane [2]). On the other hand, in systems situated between the lines $G_m^E = 1/2\, H_m^E$ and $G_m^E = H_m^E$ dipolar interactions are prevalent. This is the case of mixtures with 1,2,4-trichlorobenzene or 1-chloronaphthalene. Thus, for the 1,2,4-trichlorobenzene + hexane system, $G_m^E$/J mol$^{-1}$ = 664 [61]; $H_m^E$/J mol$^{-1}$ = 729 [13]; $(G_m^E/H_m^E)$ = 0.91, and for the 1-chloronaphthalene + hexadecane mixture, $G_m^E$/J mol$^{-1}$ = 488 [62]; $H_m^E$/J mol$^{-1}$ = 580 [14]; $(G_m^E/H_m^E)$ = 0.84. However, this class of systems shows low $\left|C_{pm}^E\right|$ values, and this suggests that dipolar interactions may be of minor importance. Note that for solutions with 1,2,4-trichlorobenzene $C_{p\text{m}}^{\text{E}}$/J mol$^{-1}$ K$^{-1}$ = $-0.91$ ($n$ = 7); $-1.89$ ($n$ = 16) [63], while for systems including 1-chloronaphthalene, $C_{p\text{m}}^{\text{E}}$/ J mol$^{-1}$ K$^{-1}$ = 0.82 ($n$ = 7) [14]; 1.34 ($n$ = 14) [63].

In order to investigate orientational effects in the present mixtures, we examine now the results obtained from the application of the Flory model (Table 4). The mean deviations, $\overline{dev}(H_\text{m}^\text{E})$, between experimental and theoretical $H_\text{m}^\text{E}$ values, defined by $\overline{dev}(H_\text{m}^\text{E}) = (1/N_\text{S})\sum dev_i(H_\text{m}^\text{E})$ where $N_S$ stands for the number of systems are: 0.021 (C$_6$H$_5$F, $N_S$ = 6) [2]; 0.031 (C$_6$H$_5$Cl, $N_S$ = 4); 0.025 (C$_6$H$_5$Br, $N_S$ = 4); 0.044 (1,2,4-trichlorobenzene, $N_S$ = 5); 0.038 (1-chloronaphthalene, $N_S$ = 6). Such results indicate that the random mixing hypothesis is rather valid for the investigated mixtures. This conclusion must be highlighted since the aromatic halogenated compounds under study are polar molecules (see above, $\mu$ (C$_6$H$_5$F) = 1.7 D [64]). The similar deviations (Table S2) obtained for mixtures with 1,2,4-trimethylbenzene $\overline{dev}(H_\text{m}^\text{E})$ = 0.030, $N_S$ = 5), or with 1-methylnaphthalene ($\overline{dev}(H_\text{m}^\text{E})$ = 0.033, $N_S$ = 5), both aromatic compounds of low dipole moment reinforce our conclusion. At this point, the "anomalous" situation of the mixtures involving 1,2,4-trichlorobenzene, or 1-chloronaphthalene within the diagram $G_\text{m}^\text{E}$ vs. $H_\text{m}^\text{E}$ remains still without explication. Below, we try to solve this problem

A number of systems show structural effects as the different signs of the excess functions $H_\text{m}^\text{E}$ (positive) and $V_\text{m}^\text{E}$ (negative) reveal (Table 8, Figures 5b and 6a). For example, $V_\text{m}^\text{E}$ ($n$ = 7)/cm$^3$mol$^{-1}$ = $-0.243$ (C$_6$H$_5$Cl) [65]; $-0.361$ (C$_6$H$_5$Br) [66]; $-0.629$ (1,2,4-trichlorobenzene) [63]; $-1.263$ (1-chloronaphthalene) [14]. The large and negative $V_\text{m}^\text{E}$ ($n$ = 7) values for systems with 1,2,4-trichlorobenzene or 1-chloronaphthalene are remarkable. Such structural effects can be due, at least in part, to free volume effects. Note the large differences between the isobaric thermal expansion coefficients of the mixture compounds, particularly for the systems containing the shorter $n$-alkanes (Figure 5a). In addition, the negative $\frac{\Delta V_\text{m}^\text{E}}{\Delta T}$ values ($-4.03$ 10$^{-3}$ cm$^3$mol$^{-1}$ K$^{-1}$ for the C$_6$H$_5$Cl + heptane mixture [67]) also indicate the existence of structural effects in the studied solutions [16,18,68].

*5.1 Mixtures with a given alkane*

For systems with, say, heptane, $H_\text{m}^\text{E}$/J mol$^{-1}$= 672 (C$_6$H$_5$Cl) [11]; 795 (C$_6$H$_5$Br) [12]; 711 (1,2,4-trichlorobenzene [13]); 656 (1-chloronaphthalene) [14]. The partial enthalpies at infinite dilution of the first compound (AHC), $H_{\text{m},1}^{\text{E},\infty}$, can be determined from $H_\text{m}^\text{E}$ measurements for the same systems over the entire composition range and we have $H_{m,1}^{E,\infty}$/kJ mol$^{-1}$ = 2.86 (C$_6$H$_5$Cl) [11] < 3.09 (C$_6$H$_5$Br) [12] < 3.35 (1-chloronaphthalene) [14] $\approx$ 3.52 (1,2,4-trichlorobenzene) [13]. The latter results suggest that interactions between like molecules are stronger in mixtures involving 1,2,4-trichlorobenzene or 1-chloronaphthalene. However, structural effects may be largely important in these systems and an analysis about the strength of the mentioned interactions should be conducted in terms of the excess molar internal energies at constant volume, $U_{Vm}^{E}$. This quantity can be determined from the equation [69]:

$$U_{Vm}^{E} = H_m^E - T\frac{\alpha_p}{\kappa_T}V_m^E \tag{13}$$

where $\alpha_p$ and $\kappa_T$ are, respectively, the isobaric thermal expansion coefficient and the coefficient of isothermal compressibility of the system under consideration. The $T\frac{\alpha_p}{\kappa_T}V_m^E$ term is defined as the contribution from the equation of state (eos) term to $H_m^E$. The $\alpha_p$ and $\kappa_T$ values have been calculated here assuming ideal behavior for the mixtures ($M^{id} = \phi_1 M_1 + \phi_2 M_2$; with $M_i = \alpha_{pi}$, or $\kappa_{T_i}$ and $\phi_i = x_i V_{m,i}/(x_1 V_{m,1} + x_2 V_{m,2})$). For pure compounds, the values used for $\alpha_{pi}$ and $\kappa_{Ti}$ are the same that in Flory calculations. In the case of heptane systems, we have $U_{Vm}^{E}$/J mol$^{-1}$ = 742 (C$_6$H$_5$Cl) < 901 (C$_6$H$_5$Br) $\simeq$ 900 (1,2,4-trichlorobenzene) < 1031 (1-chloronaphthalene) (Table 8, Figure 6b) and $U_{Vm,1}^{E,\infty}$/kJ mol$^{-1}$ = 2.91 (C$_6$H$_5$Cl) < 3.34 (C$_6$H$_5$Br) < 4.23 (1,2,4-trichlorobenzene) < 4.90 (1-chloronaphthalene). That is, interactions between like molecules of AHC become stronger in this sequence. It is pertinent to keep in mind the data corresponding to the benzene or C$_6$H$_5$F + heptane systems. Thus, we have in the case of the C$_6$H$_5$F mixture, $H_m^E$/J mol$^{-1}$ = 888, $H_{m,1}^{E,\infty}$/ kJ mol$^{-1}$ = 3.48 [60]; $U_{Vm}^{E}$/ J mol$^{-1}$ = 765 and $U_{Vm,1}^{E,\infty}$/ kJ mol$^{-1}$ = 2.9 [2], and for the benzene solution, $H_m^E$/ J mol$^{-1}$ = 931, $H_{m,1}^{E,\infty}$/ kJ mol$^{-1}$ = 3.2 [70]; $U_{Vm}^{E}$/ J mol$^{-1}$= 768 and $U_{Vm,1}^{E,\infty}$/kJ mol$^{-1}$ = 2.7 [2]. This allows to conclude that C$_6$H$_5$X - C$_6$H$_5$X interactions become stronger in the sequence X = H $\approx$ F $\approx$ Cl < Br. The similarity of the $U_{Vm}^{E}$ data for systems with C$_6$H$_6$, C$_6$H$_5$F, or C$_6$H$_5$Cl remarks again the relevance of dispersive interactions in these solutions.

The values of the molar excess free Helmholtz energy ($F_{Vm}^{E}$) can be obtained from [69,71]:

$$F_{Vm}^{E} = G_{m}^{E} - \frac{(V_{m}^{E})^{2}}{2V_{m}\kappa_{T}} \qquad (14)$$

where $V_m$ is the molar volume of the mixture. Usually, the $F_{Vm}^{E}$ and $G_m^{E}$ values are very close due to the presence of the quadratic term in $V_m^{E}$ [69,71]. For example, assuming again also ideal behavior for $\kappa_T$, the term $\frac{(V_m^{E})^2}{2V_m\kappa_T}$ is (10.9 and 0.5) J mol$^{-1}$ for the 1-chloronaphthalene + hexane, or + hexadecane mixtures, respectively. That is, we can calculate the ratios ($F_{Vm}^{E}/U_{Vm}^{E}$) for the mixtures cited above and we obtain: 0.60 ($C_6H_5Cl$ + heptane); 0.53 ($C_6H_5Br$ + heptane); 0.74 (1,2,4-trichlorobenzene + hexane); 0.69 (1-chloronaphthalene + hexadecane). It is also possible to determine the $TS_{Vm}^{E}$ values, which are, in the same order: (296, 423, 236, and 224) J mol$^{-1}$. All these results newly confirm that dispersive interactions are prevalent. It is to be noted that the two latter $TS_{Vm}^{E}$ values largely differ from the corresponding $TS_m^{E}$ values (58 and 92) J mol$^{-1}$, respectively, see above) and underline the relevance of structural effects, which overestimate, in some extent, the ($G_m^{E}/H_m^{E}$) ratios. Structural effects become more important in the sequence $C_6H_5Cl$ < $C_6H_5Br$ < 1,2,4-trichlorobenzene < 1-chloronaphthalene. The opposite variation is encountered for $V_m^{E}$ or $(\alpha_{p1} - \alpha_{p2})$ values in systems with a given $n$-alkane (Figure 5).

The replacement of hexane by cyclohexane in mixtures with $C_6H_5X$ (X = Cl, Br) does not modify the type of interactions present in the solutions (Table 4). However, there is a change in the sign of the $V_m^{E}$, in such way that both $H_m^{E}$ and $V_m^{E}$ are positive (0.281 cm$^3$ mol$^{-1}$ for the $C_6H_5Br$ system [72], Table 8), i.e., the interactional contribution to the latter excess function is dominant. This leads to for mixtures containing a given $C_6H_5X$, $H_m^{E}(n = 6)$ < $H_m^{E}(C_6H_{12})$, while $U_{Vm}^{E}$ ($n$ = 6) > $U_{Vm}^{E}$ ($C_6H_{12}$) (Table 8). Structural effects are still present since, e.g., $\frac{\Delta V_m^{E}}{\Delta T}$ = $-1.310^{-3}$ cm$^3$mol$^{-1}$ K$^{-1}$ for the $C_6H_5Br$ + $C_6H_{12}$ solution [72]. The 1,2,4-trichlorobenzene + cyclohexane mixture behaves similarly to the system with hexane and $V_m^{E}$ (= $-0.923$ cm$^3$mol$^{-1}$) and $\frac{\Delta V_m^{E}}{\Delta T}$ ( = $-0.011$ cm$^3$mol$^{-1}$ K$^{-1}$) are negative [72]. That is, the function $V_m^{E}$ is mainly determined by structural effects.

*5.2 Mixtures with a given aromatic halogenated compound*

All the systems have some common features. (i) One of the mixture component is a flat molecule. (ii) Dispersive interactions are dominant. (iii) Solutions with shorter $n$-alkanes show larger structural effects. (iv) $U_{Vm}^{E}$ values decrease when $n$ is increased (Table 8, Figure 6b). This set of features defines the Wilhelm's effect. We note that the $U_{Vm}^{E}$ variation with $n$ becomes steeper in the order: $C_6H_5Cl < C_6H_5Br <$ 1-chloronaphthalene $<$ 1,2,4-trichlorobenzene (Table 8 and Figure 6b). That is, the Wilhelm's effect is stronger in mixtures containing any of the two latter aromatic halogenated compounds, conspicuously the systems where structural effects are more relevant.

It is important to underline that benzene or $C_6H_5F$ + $n$-alkane mixtures are also characterized by the features (i) and (ii). However, the corresponding $V_m^E$ values are positive and increase with $n$ (Figure 5b) and the same occurs for the $U_{Vm}^{E}$ results [2] (Figure 6b). It is well-known that the Patterson's effect is present in solutions with benzene [1,4], and we have recently demonstrated that the mentioned effect does not exist in systems with $C_6H_5F$ [2]. On the other hand, the Patterson's effect becomes weaker when $CH_3$ groups are progressively attached to the aromatic ring. Thus, it is weaker in systems with toluene than in benzene solutions [73], while in mixtures with 1,2,4-trimethylbenzene, the Wilhelm's effect is encountered (see below).

In order to investigate the physical meaning of the $U_{Vm}^{E}(n)$ variation in the selected solutions, we pay now attention to systems formed by $C_6H_5Cl$ or 1-chloronaphthalene and a branched alkane (br-$C_n$): 2,2-dimethylbutane (br-$C_6$), 2,2,4-trimethylpentane (br-$C_8$), 2,2,4,6,6-pentamethylheptane (br-$C_{12}$), or 2,2,4,4,6,8,8-heptamethylnonane (br-$C_{16}$). Firstly, it is interesting to note that ($H_m^E$/J mol$^{-1}$) values weakly depend on the alkane in 1-chloronaphthalene systems: 942 (br-$C_6$); 928 (br-$C_8$); 1040 (br-$C_{12}$); 1003 (br-$C_{16}$) [16]. In reference [18], only values at equimolar composition for mixtures with $C_6H_5Cl$ are reported: (744 (br-$C_6$); 788 (br-$C_8$), 850 (br-$C_{12}$); 855 (br-$C_{16}$)) J mol$^{-1}$. This type of systems is also characterized by dispersive interactions and by large structural effects. The application of the Flory model to 1-chloronaphthalene mixtures yields $\overline{dev}(H_m^E) = 0.047$ ($N_S = 4$). On the other hand, $V_m^E$ (br-$C_6$)/cm$^3$mol$^{-1}$ $= -0.627$ ($C_6H_5Cl$) [18] and $-1.731$ (1-chloronaphthalene) [16]. Structural effects are particularly relevant for the systems with smaller alkanes due to their large $|\alpha_{p1} - \alpha_{p2}|$ differences [16,18]. Newly, we found that $U_{Vm}^{E}$ decreases when $n$ is increased. Thus, $U_{Vm}^{E}$(1-chloronapthalene)/J mol$^{-1}$ = 1397 (br-$C_6$); 1296 (br-$C_8$); 1243 (br-$C_{12}$); 1091 (br-$C_{16}$); and $U_{Vm}^{E}$($C_6H_5Cl$)/J mol$^{-1}$ = 925 (br-$C_6$); 875 (br-$C_8$); 828 (br-$C_{12}$); 813 (br-$C_{16}$). That is, the observed $U_{Vm}^{E}(n)$ variation is independent of the alkane shape, and may be due to larger alkanes break a less number of AHC-AHC interactions. Dynamic viscosity ($\eta$) data support this assumption. Thus, the magnitude $\Delta \eta (= x_1 \eta_1 + x_2 \eta_2)$ becomes less negative when $n$ increases in systems with 1,2,4-trichlorobenzene or 1-chloronaphthalene. In the former case, $\Delta \eta$ /mPa s = −0.30 ($n$ = 10); −0.08 ($n$ = 14) [53] and in mixtures with 1-chloronaphthalene, $\Delta \eta$ /mPa s =−1.29 ($n$ = 6); −1.19 ($n$ = 8), −0.98 ($n$ = 12) [74]. In contrast, benzene mixtures, where Patterson's effect exists, are characterized by decreasing ($\Delta \eta$ /mPa s) values when $n$ is increased: −0.06 ($n$ = 10); −0.09 ($n$ = 12); −0.14 ($n$ = 14); −0.24 ($n$ = 16) [75]. The fact that, for systems which differ merely in the shape of the alkane (same $n$), $U_{Vm}^{E}$ ($n$-C$_n$) < $U_{Vm}^{E}$ (br-C$_n$) may be due to branched alkanes are better breakers of the AHC-AHC interactions than $n$-alkanes. The next step is to examine the values of the excess molar heat capacities at constant volume, $C_{Vm}^{E}$, determined from the well-known equation [76]:

$$C_{Vm}^{E} = C_p \frac{\kappa_S}{\kappa_T} - C_p^{id} \frac{\kappa_S^{id}}{\kappa_T^{id}} = C_{pm}^{E} - \left[ \frac{TV_m \alpha_p^2}{\kappa_T} - \frac{TV_m^{id} \alpha_p^{id2}}{\kappa_T^{id}} \right] \qquad [15]$$

where $\kappa_S$ is the isentropic compressibility of the mixture. The $\left[ \frac{TV_m \alpha_p^2}{\kappa_T} - \frac{TV_m^{id} \alpha_p^{id2}}{\kappa_T^{id}} \right]$ term can be considered as the contribution from the eos term to $C_{pm}^{E}$. From the values obtained, some important statements can be provided. (i) $C_{Vm}^{E}$ values are negative (Figure 7) and this confirms again that dispersive interactions are dominant. (ii) Both $C_{Vm}^{E}$ and $n$ increase in line in systems involving branched alkanes (Figure 7), which reveals that a lower number of AHC-AHC interactions are broken by the larger branched alkanes. (iii) The same trend is observed for systems with $n$-alkanes up to $n$ = 10. However, from $n$ = 12, $C_{Vm}^{E}$ decreases (Figure 7). This may be indicative of that the short orientational order existing in longer $n$-alkanes is broken by the AHC molecules. Note that for systems with 1-chloronaphthalene, $U_{Vm}^{E}$ (dodecane) − $U_{Vm}^{E}$ (hexadecane) = 116 J mol$^{-1}$ (Table 8) while $U_{Vm}^{E}$ (br-C$_{12}$) − $U_{Vm}^{E}$ (br-C$_{16}$) = 152 J mol$^{-1}$. Solutions with C$_6$H$_5$Cl deserve a specific comment, since $H_m^{E}$ data reported in ref. [18] differ from those given in references [11,77]. Particularly, for the system with $n$ = 14, $H_m^{E}$/J mol$^{-1}$ = 755 [11], 838 [18], which is a meaningful difference in view of that the considered effects are rather subtle. Accordingly with the data from reference [18], $U_{Vm}^{E}$ decreases up to $n$ = 10 and then increases, which seems to points out to the existence of an extra endothermic contribution to $U_{Vm}^{E}$ from the disruption of the short orientational order in longer *n*-alkanes. Nevertheless, this needs experimental confirmation.

$C_{pm}^{E}$ values are then the result of an interactional contribution and of a structural contribution, which is, in fact, largely relevant and that leads to a "parabolic" dependence of $C_{pm}^{E}$ with *n* [14,63,78] (Figure 7).

On the other hand, $V_{m}^{E}$ increases with *n*, and in systems formed by C$_6$H$_5$X (X = Cl, Br) or 1,2,4-trichlorobenzene and longer *n*-alkanes becomes positive (Figure 5b, Table 8), i.e., structural effects are weaker in such solutions. Accordingly, $\frac{\Delta V_{m}^{E}}{\Delta T}$ is slightly positive in mixtures with C$_6$H$_5$Cl and $n \geq 10$ [18].

*5.3 1,2,4-Trimethylbenzene or 1-methylnaphthalene + n-alkane mixtures*

It has been already mentioned that results from the application of the Flory model show that dispersive interactions are dominant in these systems. Consequently, they are located between the lines $G_{m}^{E} = 1/3\,H_{m}^{E}$ and $G_{m}^{E} = 1/2\,H_{m}^{E}$ in the $G_{m}^{E}$ vs $H_{m}^{E}$ diagram. For the heptane mixtures, DISQUAC provides $G_{m}^{E}$ / J mol$^{-1}$ = 198 (1-methylnaphthalene) and 127 (1,2,4-trimethylbenzene). In the same order, $H_{m}^{E}$/ J mol$^{-1}$ = 759 and 276 [17], and ($G_{m}^{E}/H_{m}^{E}$) = 0.26 and 0.46. In addition, structural effects are also present, which is supported by the following values of $V_{m}^{E}$ (*n* = 7)/cm$^3$mol$^{-1}$: −0.207 (1,2,4-trimethylbenzene) [17], −1.361 (1-methylnaphthalene; *T* = 303.15 K [79]). We note that for systems with a given *n*-alkane, $H_{m}^{E}$(1,2,4-trimethylbenzene) < $H_{m}^{E}$(1-methylnaphthalene) (Table S2), which can be ascribed to the aromatic surface is smaller in the alkylbenzene. Both $H_{m}^{E}$ (Tables 4 and S2), and $U_{Vm}^{E}$ become higher when, in systems with a given *n*-alkane, 1,2,4-trimethylbenzene is replaced by 1,2,4-trichlorobenzene due to interactions between like molecules are stronger in the latter case. In fact, in heptane solutions $U_{Vm,1}^{E,\infty}$(1,2,4-trimethylbenzene)/kJ mol$^{-1}$ = 1.07 < $U_{Vm,1}^{E,\infty}$(1,2,4-trichlorobenzene) = 3.48. We have roughly estimated $U_{Vm}^{E}$ for the systems 1-methylnaphthalene + heptane using the $V_{m}^{E}$ value given above. The result (1142 J mol$^{-1}$) is higher than that for the solution with 1-chloronaphthalene and suggests that more dispersive interactions between like molecules are disrupted upon mixing in the former case. In mixtures including 1,2,4-trimethylbenzene, $H_{m}^{E}$ depends weakly on *n* (Table S2). However, $U_{Vm}^{E}$/J mol$^{-1}$ decreases when *n* is increased: 338 (*n* = 7); 232 (*n* = 10); 209 (*n* = 14) and we can conclude that the Wilhelm's effect exists in these mixtures. Regarding to solutions with 1-methylnaphthalene, the same conclusion is still valid. In fact, we have not been able to determine $U_{V_m}^E$ values due to the lack of the required volumetric data, however, the decreasing $H_m^E$ values when $n$ is increased (Table S2) together with the large structural effects present in the mixture with heptane strongly support our conclusion.

*5.4 The concentration-concentration structure factor*

All the systems show homocoordination ($S_{CC}(0) > 0.25$) (Table 7 and Figure 4). DISQUAC calculations reveal that $S_{CC}(0)$ ($n = 7$) decreases in the sequence: C$_6$H$_5$Cl ≈ C$_6$H$_5$Br < 1-chloronaphthalene < 1,2,4-trichlorobenzene, while $S_{CC}(0)$ values become very similar when 1-chloronaphthalene is replaced by 1,2,4-trichlorobenzene in mixtures with tetradecane. It is to be noted that the $S_{CC}(0)$ variation is more or less in agreement with the situation of the systems in the $G_m^E$ vs. $H_m^E$ diagram and mixtures which are upper from the line $G_m^E = H_m^E/2$ show a higher homocoordination. For mixtures containing C$_6$H$_5$Cl, $S_{CC}(0)$ decreases when $n$ is increased: 0.359 ($n = 7$) [67] > 0.267 ($n = 10$) [80] (values at 323.15 K, see Table 7). In terms of DISQUAC, this behaviour is common to all the solutions with a given AHC and may be ascribed to increasing size effects (Table 7). In the case of systems with 1,2,4-trichlorobenzene, $S_{CC}(0)^{-1}$(comb) = 0.0013 ($n = 7$); 0.4349 ($n = 14$). In addition, $S_{CC}(0)^{-1}$(int) increases in line with $n$: −2.297 ($n = 7$); −1.832 ($n = 14$), which also leads to decreased $S_{CC}(0)$ values.

In 1-chloronaphthalene mixtures, $S_{CC}(0)$ (br-C$_{16}$) > $S_{CC}(0)$ (n-C$_{16}$) [62,81]. It must be underlined that the $C_{pm}^E$ curve of the 1-chloronaphthalene + br-C$_{16}$ mixture is W-shaped [62], and that this fact together with its large $S_{CC}(0)$ value have been considered as a manifestation of non-randomness effects occurring above ~100 K of the upper critical solution temperature (UCST) [62]. $S_{CC}(0)$ also decreases when 2,2,4-trimethylpentane ($S_{CC}(0) = 0.411$) [82]) is replaced by octane ($S_{CC}(0) = 0.368$, DISQUAC result) in C$_6$H$_5$Cl systems at 303.15 K.

$S_{CC}(0)$ for 1-methylnaphthalene or 1,2,4-trimethylbenzene + heptane mixtures are 0.297 and 0.278 (DISQUAC results), respectively, which are lower than the results for the systems with 1-chloronaphthalene or 1,2,4-trichlorobenzene (Table 7), in agreement with the position of the former solutions in the $G_m^E$ vs. $H_m^E$ diagram.

*5.5 Proximity effects between the phenyl ring and the Cl atom*

In order to examine such effects, we consider the systems C$_6$H$_5$-(CH$_2$)$_u$-Cl + heptane ($u = 1,2,3$) which show common features with the corresponding solutions formed by C$_6$H$_5$Cl. Firstly, it must be stated that they are also characterized by dispersive interactions. In fact, their position in the diagram $G_m^E$ vs. $H_m^E$ and the results provided by the Flory model support such conclusion. Thus, $G_m^E$/J mol$^{-1}$ = 800 ($u$ = 1); 831 ($u$ = 2) and 658 ($u$ = 3) (DISQUAC calculations using interaction parameters from [83]); and $H_m^E$/J mol$^{-1}$= 1552 ($u$ = 1); 1429 ($u$ = 2); 1082 ($u$ = 3) [83]. Therefore, ($G_m^E / H_m^E$) = 0.51 ($u$ = 1); 0.58 ($u$ = 2); 0.61 ($u$ = 3). For the mixture with $u$ =1, the Flory model gives $dev(H_m^E)$ = 0.028. No data are available on $\alpha_p$ and $\kappa_T$ for the remainder alkyl-chlorobenzenes to complete Flory calculations. In addition, the $TS_m^E$ values are also large and positive. On the other hand, structural effects are still present in the mentioned solutions as the negative $V_m^E$ value of the benzyl chloride + octane system at 293.15 K ($-0.1066$ cm$^3$mol$^{-1}$) indicates [65]. From the enthalpic data listed above, we note that $H_m^E$ changes in the sequence: ($u$ =1) > ($u$ = 2) > ($u$ = 3) > ($u$ = 0). The corresponding ($H_{m,1}^{E,\infty}$/kJ mol$^{-1}$) values are: 2.86 ($u$ = 0) [11], 5.90 ($u$ = 1); 6.41 ($u$ = 2) and 5.97 ($u$ = 3) [83]. It seems clear that interactions between chlorobenzene molecules are weaker. We investigate now if the observed differences between the $H_m^E$ data can be ascribed to the different polarity of alkyl-chlorobenzenes. This analysis is conducted in terms of the effective dipole moments, $\bar{\mu}$, of C$_6$H$_5$-(CH$_2$)$_u$-Cl, an useful magnitude to examine the impact of polarity of bulk properties, that can be calculated using the equation [32,69,84]:

$$\bar{\mu} = \left[ \frac{\mu^2 N_A}{4\pi\varepsilon_0 V_m k_B T} \right]^{1/2} \tag{16}$$

where all the symbols have the usual meaning. The dipole moments (in D) of alkyl-chlorobenzenes are: 1.85 ($u$ = 1); 1.80 ($u$ = 2) and 1.78 ($u$ = 3) [39] and the molar volumes in the same order are: (118.37; 131.79; 148.43) cm$^3$ mol$^{-1}$ [83]. Thus, $\bar{\mu}$ = 0.598 ($u$ = 0); 0.651 ($u$ = 1); 0.600 ($u$ = 2) and 0.559 ($u$ = 3). Alkyl-chlorobenzenes with $u$ = 0, or $u$ = 2 have practically the same $\bar{\mu}$ value, while the $H_m^E$ value of the C$_6$H$_5$-(CH$_2$)$_2$-Cl + heptane system is 757 J mol$^{-1}$ larger than the value of the chlorobenzene mixture. Such large difference can be explained in terms of proximity effects between the C$_6$H$_5$- ring and the Cl atom which enhance the interactions between like molecules of alkyl-chlorobenzene. The mentioned effects are stronger when $u$ = 1 or 2, and are indeed weaker in chlorobenzene. We remark that for heptane mixtures, DISQUAC yields using interaction parameters from [83]: $S_{CC}(0)$ ($u$ = 1) = 0.514 > $S_{CC}(0)$ ($u$ = 0) = 0.380. A similar $H_m^E$ variation is encountered for C$_6$H$_5$-(CH$_2$)$_u$-CO-CH$_3$ + $n$-alkane mixtures since $H_m^E$($n$-C$_7$)/J mol$^{-1}$ = 1680 ($u$ =1) [85] > 1604 ($u$ = 2) [86] > 1492 ($u$ = 0) [85]. Interactions between like molecules become weaker in the same order, which can be explained in terms of decreasing proximity effects in the indicated sequence. It is pertinent to remark that the variation of these enthalpic data is consistent with the variation of UCSTs of these solutions. Thus, UCST($n$-C$_{10}$)/K = 301.6 ($u$ =1) > 284.5 ($u$ = 2) [87] > 277.4 ($u$ = 0) [88]. Nevertheless, it is must be pointed out that the change of proximity effects with the separation between the phenyl ring and the polar group strongly depends on the polar group involved. For example, systems with C$_6$H$_5$-(CH$_2$)$_u$-X (X = O or CHO) behave differently. $H_m^E$ ($n$-C$_7$, X = O)/J mol$^{-1}$ = 1278 ($u$ =0) > 1213 ($u$ = 1) [86] > 1148 ($u$ =2) [89]; and $H_m^E$ (X = CHO; $n$-C$_7$) /J mol$^{-1}$ = 1480 ($u$ = 1) > 1367 ($u$ = 0) > 1061 ($u$ = 2) [90]. More details can be found elsewhere [91,92].

**RA** *5.6 Aromacity effect*

This effect is investigated by comparing results between CH$_3$(CH$_2$)$_5$X or C$_6$H$_5$X (X = Cl, Br) + $n$-alkane mixtures  Systems with 1-chlorohexane or 1-bromohexane and a $n$-alkane show lower $H_m^E$ values than the corresponding mixtures with the halogenated aromatic compounds. For example, $H_m^E$/J mol$^{-1}$ = 396 (1-chlorohexane + heptane) [93]; 401 (1-bromohexane + hexane, $T$ = 303.15 K) [94]. The corresponding $G_m^E$ /J mol$^{-1}$ values, determined from DISQUAC calculations using interactions parameters from the literature [30,31] are, respectively, 206 and 204, and the ($G_m^E$ / $H_m^E$) values are, in the same order 0.52 and 0.51. In view of their position in the $G_m^E$ vs. $H_m^E$ diagram, one can conclude that dispersive interactions are also determinant in these systems. The available $H_m^E$ data at different temperatures for similar mixtures reveal that this excess function decreases when $T$ is increased. In the case of the 1-chloropentane + cyclohexane mixture, $\frac{\Delta H_m^E}{\Delta T}$ = $-1.55$ J mol$^{-1}$K$^{-1}$ [95]. The application of the Flory model to the systems 1-chlorohexane + hexane, or + heptane, or + octane [93] and to 1-bromohexane + hexane [94] yields $\overline{dev(H_m^E)}$ = 0.046. These results confirm that orientational effects are rather weak in the considered solutions. On the other hand, $V_m^E$ decreases when $T$ increases. Thus, for the 1-chlorohexane + heptane system, $\frac{\Delta V_m^E}{\Delta T}$ = $-5.1$ 10$^{-4}$ cm$^3$ mol$^{-1}$ K$^{-1}$ [96,97] and this means that structural effects are here also relevant.

The dipole moments of 1-chlorohexane and 1-bromohexane are 1.94 D and 1.99 D [39], respectively, and their effective dipole moments, in the same order, are 0.633 and 0.641 (results determined using $V_m$ from [96,98]). These $\bar{\mu}$ values are slightly higher to those of the halogenated aromatic compounds (0.598 and 0.585 for C$_6$H$_5$Cl and C$_6$H$_5$Br, respectively). It seems that the $H_m^E$ differences between $n$-alkane systems containing haloalkanes or C$_6$H$_5$X (X=

Cl, Br) molecules can not be explained in terms of the different polarity of these compounds. The replacement of *n*-alkane by benzene in systems with a given haloalkane leads to decreased $H_m^E$ values [99,100]. For example, $H_m^E$ (1-chlorohexane + benzene) = 77 J mol$^{-1}$ [99]. Such behaviour can be ascribed to the interactions between unlike molecules created upon mixing. Thus, the larger $H_m^E$ values for C$_6$H$_5$X (X = Cl, Br) + *n*-alkane mixtures are due to the presence of the phenyl ring in the polar compound (proximity effect) enhances interactions between like molecules. For the 1-chlorohexane + heptane mixture, $V_m^E$ is negligible (0.0012 cm$^3$mol$^{-1}$ [96]) and $H_m^E \simeq U_{Vm}^E$ and $H_{m,1}^{E,\infty} \simeq U_{Vm,1}^{E,\infty} = 2.31$ kJ mol$^{-1}$ [93]. This value is lower than that for the chlorobenzene solution, indicating that interactions between 1-chlorohexane molecules are weaker. Finally, it is important to note that homocoordination is higher in systems with aromatic halogenated compounds as the following DISQUAC values of $S_{CC}(0)$ reveal: 0.299 (1-chlorohexane + heptane); 0.305 (1-bromohexane + heptane) [30,31].

5.7 *The interaction parameters.*

The $C_{ag,1}^{QUAC}$ coefficients (l =1,2,3) are the same for systems with C$_6$H$_5$Cl, C$_6$H$_5$Br or 1-chloronaphthalene, while they are different for mixtures including 1,2,4-trichlorobenzene. That is, in terms of DISQUAC, the studied systems are more sensitive to the increasing of the number of halogen atoms attached to the aromatic ring than to the specific halogen atom involved in the molecule, or to the increasing of the aromatic surface. We underline that $C_{ag,1}^{QUAC}$ coefficients (l =1,3) are also used in solutions with C$_6$H$_5$F [2].

In absence of the needed data to calculate $U_{Vm}^E$ and $C_{Vm}^E$ values, the model applications may be useful to detect the existence of an anomalous effect. In the case of the Flory model, the $X_{12}$ parameter decreases when *n* is increased for most of the systems considered along the present work (Tables 1 and S1). This is indeed an unexpected behaviour in solutions with C$_6$H$_5$Cl or C$_6$H$_5$Br, where $H_m^E$ and *n* increase in line, and can be linked to the presence of the Wilhelm's effect. For mixtures containing 1,2,4-trimethylbenzene, $X_{12}$ is nearly constant from *n* = 12, but the value is lower than for the system with *n* = 7. In terms of DISQUAC, the matter is somewhat difficult. We note that if the mixture is built by only one contact (systems with 1-methylnaphthalene or 1,2,4-trimethylbenzene), the variation of the $C_{ab,2}^{DIS}$ coefficient is similar to that of $X_{12}$ (Table S1). If the solution is built by several contacts, the variation of $C_{ag,2}^{DIS}$ with *n* depends on the $C_{ab,2}^{DIS}$ values. Thus, the $C_{ag,2}^{DIS}$ coefficients decrease when *n* is increased in mixtures with C$_6$H$_5$Cl or C$_6$H$_5$Br, increase in systems with 1-chloronaphthalene and hardly depends on n in solutions with 1,2,4-trichlorobenzene (Table 1). In view of these considerations, the application of the Flory model is more suitable in order to investigate the possible existence of the Wilhelm's effect when the required experimental data are not available.

Finally, we remember that $H_m^E$ data are correctly represented when interaction parameters dependent on the *n*-alkane are used in systems with benzene or toluene, which indicates the existence of the Patterson's effect. In contrast, the effect is not observed in systems with $C_6H_5F$, since the $H_m^E$ data are well described by means of enthalpic parameters independent on the *n*-alkane [2].

## 6. Conclusions

Binary mixtures formed by $C_6H_5X$ (X = Cl, Br), or 1-chloronaphthalene, or 1,2,4-trichlorobenzene, or 1-methylnaphthalene or 1,2,4-trimethylbenzene and alkane have been investigated using the DISQUAC, Flory and $S_{CC}(0)$ models, and by means of experimental data, including excess functions at constant volume. Systems with AHC are characterized by dispersive interactions, large structural effects when smaller alkanes are involved and decreasing $U_{Vm}^E$ values when *n* is increased. This variation is observed in systems with linear or branched alkanes and suggests that larger alkanes break a lower number of AHC-AHC interactions. Accordingly, $C_{Vm}^E$ and *n* increases in line in solutions with branched alkanes. The same variation is observed in systems with *n*-alkanes up to *n* = 10, and then $C_{Vm}^E$ decreases, which may be ascribed to the breaking of the local orientational order of longer *n*-alkanes. Aromacity effects leads to stronger interactions between halogenated molecules. Proximity effects in $C_6H_5$-$(CH_2)_u$-Cl + heptane mixtures are stronger when *u* =2,3 and are weaker in the solution with chlorobenzene.

**Funding**

This work was supported by Consejería de Educación de Castilla y León, under Project VA100G19 (Apoyo a GIR, BDNS: 425389).

**TABLE 1**

Dispersive (DIS) and quasichemical (QUAC) interchange coefficients, $C_{sg,l}^{DIS}$ and $C_{sg,l}^{QUAC}$, for (s,g) contacts[a] in aromatic halogenated compound + alkane mixtures ($l = 1$, Gibbs energy; $l = 2$, enthalpy; $l = 3$, heat capacity). The interaction parameter, $X_{12}$, of the Flory model is also given.

| Alkane | DISQUAC interchange coefficients[b] | | | | | | Flory parameter |
|---|---|---|---|---|---|---|---|
| | $C_{sg,1}^{DIS}$ | $C_{sg,2}^{DIS}$ | $C_{sg,3}^{DIS}$ | $C_{sg,1}^{QUAC}$ | $C_{sg,2}^{QUAC}$ | $C_{sg,3}^{QUAC}$ | $X_{12}$/J cm$^{-3}$ |
| $C_6H_5Cl$ (s = a) | | | | | | | |
| $n$-C$_6$ | −1.75 | −1.57 | 0.1 | 2 | 1.75 | 0.2 | 30 |
| $n$-C$_7$ | −1.75 | −1.62 | 0.2 | 2 | 1.75 | 0.2 | 28.77 |
| $n$-C$_8$ | −1.75 | −1.65[c] | 0.23 | 2 | 1.75 | 0.2 | |
| $n$-C$_9$ | −1.75 | −1.70 | 0.30[e] | 2 | 1.75 | 0.2 | 28.22 |
| $n$-C$_{10}$ | −1.86 | −1.73[c] | 0.35 | 2 | 1.75 | 0.2 | |
| $n$-C$_{12}$ | −1.86 | −1.80[c] | 0.5 | 2 | 1.75 | 0.2 | |
| $n$-C$_{14}$ | −1.86 | −1.88 | 0.8 | 2 | 1.75 | 0.2 | 27.28 |
| $n$-C$_{16}$ | −1.86 | −1.88[c] | 1.2 | 2 | 1.75 | 0.2 | |
| $C_6H_5Cl$ (s = c) | | | | | | | |
| $C_6H_{12}$ | −1.79 | −1.54 | −0.1 | 2 | 1.75 | 0.2 | 32.71 |
| $C_6H_5Br$ (s = a) | | | | | | | |
| $n$-C$_6$ | −1.69 | −1.33 | 0.1[c] | 2 | 1.75 | 0.2 | 34.40 |
| $n$-C$_7$ | −1.69 | −1.33 | 0.1[c] | 2 | 1.75 | 0.2 | 34.40 |
| $n$-C$_9$ | −1.69 | −1.45 | 0.3[c] | 2 | 1.75 | 0.2 | 31.19 |
| $n$-C$_{14}$ | −1.69 | −1.68 | 0.9[c] | 2 | 1.75 | 0.2 | 28.78 |
| $C_6H_5Br$ (s = c) | | | | | | | |
| $C_6H_{12}$ | −1.6 | −1.36 | 0.8 | 2 | 1.75 | 0.2 | 34.08 |
| 1-chloronaphthalene (s = a) | | | | | | | |
| $n$-C$_6$ | −1.05[c] | −2.15 | −0.9 | 2 | 1.75 | 0.2 | 27.28 |
| $n$-C$_7$ | −1.05[c] | −2.15 | −0.25 | 2 | 1.75 | 0.2 | 24.89 |
| $n$-C$_8$ | −1.05[c] | −2.08 | −0.05 | 2 | 1.75 | 0.2 | 23.63 |

TABLE 1 (continued)

| | | | | | | | |
|---|---|---|---|---|---|---|---|
| $n$-C$_{10}$ | −1.3 | −1.90 | 0.05 | 2 | 1.75 | 0.2 | 21.33 |
| $n$-C$_{12}$ | −1.3 | −1.79 | 0.1 | 2 | 1.75 | 0.2 | 19.15 |
| $n$-C$_{14}$ | −1.3 | −1.79$^c$ | −0.15 | 2 | 1.75 | 0.2 | |
| $n$-C$_{16}$ | −1.3 | −1.79$^c$ | −0.3 | 2 | 1.75 | 0.2 | 16.19 |
| 1,2,4-trichlorobenzene (s = a) | | | | | | | |
| $n$-C$_6$ | −0.182 | −2.65 | −1.62 | 0.5 | 2.8 | 0.2 | 29.89 |
| $n$-C$_7$ | −0.182 | −2.65 | −1.62 | 0.5 | 2.8 | 0.2 | 27.28 |
| $n$-C$_9$ | −0.182 | −2.65 | −1.62 | 0.5 | 2.8 | 0.2 | 23.64 |
| $n$-C$_{10}$ | −0.35 | −2.65 | −1.62 | 0.5 | 2.8 | 0.2 | 22.97 |
| $n$-C$_{12}$ | −0.35 | −2.65$^c$ | −1.26 | 0.5 | 2.8 | 0.2 | |
| $n$-C$_{14}$ | −0.35 | −2.7 | −1.26$^c$ | 0.5 | 2.8 | 0.2 | 18.44 |
| $n$-C$_{16}$ | −0.35 | −2.73$^c$ | −1.26 | 0.5 | 2.8 | 0.2 | |

$^a$type s = a, aliphatic, type s = c, cyclic; type g, Cl or Br; $^b$ $C_{sg,1}^{DIS/QUAC}$ coefficients determined assuming that the interchange coefficients for the (b,g) contacts are 0 (see text); $^c$estimated value

**TABLE 2**

Physical properties of pure compounds at 298.15 K and $p = 0.1013$ MPa: $V_m$, molar volume; $\alpha_p$, isobaric thermal expansion coefficient; $\kappa_T$, isothermal compressibility. The reduction parameters in the Flory model are also included: $V_m^*$, reduction molar volume; $P^*$, reduction pressure.

| Compound | $V_m$/cm$^3$·mol$^{-1}$ | $\alpha_p$/ 10$^{-3}$K$^{-1}$ | $\kappa_T$/ TPa$^{-1}$ | $V_m^*$/cm$^3$·mol$^{-1}$ | $P^*$/MPa |
|---|---|---|---|---|---|
| chlorobenzene | 102.24[a] | 0.99[a] | 771[a] | 82.08 | 594 |
| bromobenzene | 105.50[a] | 0.933[a] | 668[b] | 85.51 | 633.9 |
| 1-chloronaphthalene | 136.77[c] | 0.702[c] | 488[c] | 115.59 | 600.5 |
| 1,2,4-trichlorobenzene | 125.28[d] | 0.826[d] | 558[d,e] | 103.46 | 647.1 |
| 1-methylnaphthalene | 139.69[f] | 0.73[f] | 535.2[f] | 117.42 | 575.7 |
| 1,2,4-trimethylbenzene | 137.89[g] | 0.926[g] | 792[g,h] | 111.89 | 529.4 |

[a][101]; [b][102]; [c][16]; [d][103]; [e][104]; [f][105]; [g][106]; [h][17]

**TABLE 3**

Molar excess Gibbs energy, $G_m^E$, at equimolar composition and temperature $T$, for aromatic halogenated compound mixtures. Comparison of experimental results (Exp.) with DISQUAC (DQ) calculations using the interaction parameters from Table 1. $N$ is the number of data points and $\sigma_r(P)$ the relative standard deviation for pressures (equation 10).

| System | $T$/K | $N$ | $G_m^E$ /J·mol$^{-1}$ | | $\sigma_r(P)$ | | Ref. |
|---|---|---|---|---|---|---|---|
| | | | Exp. | DQ | Exp. | DQ | |
| Chlorobenzene + $n$-C$_6$ | 338.2 | 12 | 459 | 439 | 0.005 | 0.010 | 107 |
| Chlorobenzene + $n$-C$_7$ | 323.15 | 16 | 415 | 428 | 0.002 | 0.006 | 67 |
| | 353.15 | 19 | 392 | 409 | 0.003 | 0.007 | 67 |
| | 373.15 | 20 | 383 | 397 | 0.002 | 0.007 | 67 |
| Chlorobenzene + $n$-C$_{10}$ | 323.15 | 10 | 275 | 311 | 0.011 | 0.021 | 80 |
| | 348.15 | 12 | 271 | 284 | 0.006 | 0.008 | 80 |
| | 373.15 | 15 | 290 | 261 | 0.013 | 0.023 | 80 |
| Chlorobenzene + C$_6$H$_{12}$ | 348.15 | 26 | 329 | 330 | 0.003 | 0.010 | 108 |
| Bromobenzene + $n$-C$_7$ | 382 | | 397 | 401 | | | 109 |
| Bromobenzene + C$_6$H$_{12}$ | 298.15 | 29 | 452 | 448 | 0.011 | 0.013 | 110 |
| | 308.15 | 30 | 438 | 438 | 0.001 | 0.0004 | 110 |
| 1,2,4-trichlorobenzene + $n$-C$_6$ | 283.15 | 16 | 634 | 681 | 0.045 | 0.150 | 110 |
| | 293.15 | 17 | 680 | 679 | 0.014 | 0.017 | 110 |
| | 303.15 | 17 | 669 | 677 | 0.024 | 0.020 | 110 |
| 1-chloronaphthalene + $n$-C$_{16}$ | 298.15 | | 488 | 480 | | | 62 |

**TABLE 4**

Molar excess enthalpies, $H_m^E$, at equimolar composition, temperature $T$ and $p = 0.1013$ MPa for aromatic halogenated compound (1) + alkane (2) mixtures. Comparison of experimental results with DISQUAC (DQ) and Flory calculations using the interaction parameters from Table 1. $N$ is the number of data points and $dev(H_m^E)$ the relative deviation for $H_m^E$ (equation 11).

| alkane | T/K | N | $H_m^E$/J·mol$^{-1}$ | | $dev(H_m^E)$ | | | Ref. |
|---|---|---|---|---|---|---|---|---|
| | | | Exp. | DQ | Exp. | DQ | Flory | |
| Chlorobenzene | | | | | | | | |
| n-C$_6$ | 298.15 | 18 | 662 | 661 | 0.007 | 0.011 | 0.022 | 77 |
| n-C$_7$ | 288.15 | 20 | 625 | 682 | 0.006 | 0.070 | | 67 |
| | 298.15 | 10 | 672 | 673 | 0.005 | 0.021 | 0.039 | 11 |
| | | 12 | 696 | | 0.013 | 0.027 | | 83 |
| | | 10 | 671 | | 0.005 | 0.021 | | 111 |
| | | 8 | 643 | | 0.026 | 0.112 | | 112 |
| | 308.15 | 20 | 607 | 664 | 0.011 | 0.079 | | 67 |
| | 318.15 | 7 | 610 | 655 | 0.010 | 0.079 | | 112 |
| n-C$_9$ | 298.15 | 10 | 712 | 703 | 0.007 | 0.022 | 0.037 | 11 |
| n-C$_{14}$ | 298.15 | 11 | 755 | 749 | 0.007 | 0.023 | 0.025 | 11 |
| C$_6$H$_{12}$ | 298.15 | 20 | 699 | 688 | 0.013 | 0.020 | 0.046 | 113 |
| | | 12 | 753 | | 0.015 | 0.072 | | 83 |
| | | 7 | 698 | | 0.005 | 0.017 | | 114 |
| | 308.65 | 9 | 665 | 662 | 0.018 | 0.042 | | 115 |
| | 318.15 | 9 | 645 | 636 | 0.015 | 0.048 | | 115 |
| | 328.15 | 9 | 622 | 609 | 0.012 | 0.051 | | 115 |
| Bromobenzene | | | | | | | | |
| n-C$_6$ | 298.15 | 13 | 772 | 753 | 0.006 | 0.023 | 0.021 | 12 |
| n-C$_7$ | 298.15 | 13 | 795 | 796 | 0.004 | 0.018 | 0.022 | 12 |
| n-C$_9$ | 298.15 | 10 | 805 | 811 | 0.002 | 0.013 | 0.021 | 12 |
| n-C$_{14}$ | 298.15 | 10 | 824 | 830 | 0.007 | 0.031 | 0.037 | 12 |
| C$_6$H$_{12}$ | 298.15 | 10 | 740 | 741 | 0.004 | 0.023 | 0.032 | 12 |
| 1-Chloronaphthalene | | | | | | | | |
| n-C$_6$ | 298.15 | 14 | 665 | 665 | 0.009 | 0014 | 0.042 | 14 |
| n-C$_7$ | 298.15 | 17 | 656 | 661 | 0.007 | 0.012 | 0.046 | 14 |
| n-C$_8$ | 298.15 | 19 | 658 | 658 | 0.007 | 0.020 | 0.043 | 14 |
| n-C$_{10}$ | 298.15 | 19 | 654 | 652 | 0.010 | 0.017 | 0.031 | 14 |
| n-C$_{12}$ | 298.15 | 22 | 630 | 644 | 0.006 | 0021 | 0.018 | 14 |

TABLE 4 (continued)

| | | | | | | | | |
|---|---|---|---|---|---|---|---|---|
| $n$-C$_{16}$ | 298.15 | 15 | 580 | 580 | 0.006 | 0.057 | 0.047 | 14 |
| 1,2,4-trichlorobenzene | | | | | | | | |
| $n$-C$_6$ | 298.15 | 12 | 729 | 738 | 0.007 | 0.013 | 0.038 | 13 |
| $n$-C$_7$ | 298.15 | 12 | 711 | 728 | 0.007 | 0.021 | 0.041 | 13 |
| $n$-C$_9$ | 298.15 | 12 | 680 | 693 | 0.007 | 0.014 | 0.043 | 13 |
| $n$-C$_{10}$ | 298.15 | 11 | 681 | 674 | 0.008 | 0.018 | 0.045 | 13 |
| $n$-C$_{14}$ | 298.15 | 12 | 608 | 628 | 0.007 | 0.023 | 0.055 | 13 |

**TABLE 5**

Results on solid-liquid equilibria for 1,2,4-trichlorobenzene (1) + $n$-alkane (2) systems from the application of the IDEAL and DISQUAC models: absolute mean deviations, $\Delta(T)$ [a], and relative standard deviations, $\sigma_r(T)$ [b], for temperature and coordinates of the eutectic points: composition, $x_{1eu}$, and temperature, $T_{eu}$. Experimental data are taken from reference [53].

| $n$-alkane | $\Delta(T)$/K | | $\sigma_r(T)$ | |
|---|---|---|---|---|
| | IDEAL | DQ | IDEAL | DQ |
| $n$-C$_{10}$ | 7.4 | 0.68 | 0.028 | 0.002 |
| $n$-C$_{14}$ | 1.5 | 0.52 | 0.009 | 0.003 |

| | Eutectic coordinates | | | | | |
|---|---|---|---|---|---|---|
| | $x_{1eu}$ | | | $T_{eu}$/K | | |
| | Exp. | IDEAL | DQ | Exp. | IDEAL | DQ |
| $n$-C$_{10}$ | 0.100 | 0.212 | 0.096 | 242 | 239.4 | 241.9 |
| $n$-C$_{14}$ | 0.450 | 0.539 | 0.426 | 272.1 | 268.2 | 272.6 |

[a] $\Delta(T)/K = \dfrac{1}{N}\sum\left|T_{exp} - T_{calc}\right|$; [b] $\sigma_r(T) = \left\{\dfrac{1}{N}\sum\left[\dfrac{T_{exp} - T_{calc}}{T_{exp}}\right]^2\right\}^{1/2}$

**TABLE 6**

Isobaric molar excess heat capacities, $C_{pm}^{E}$, of aromatic halogenated compound (1) + $n$-alkane (2) mixtures at 298.15 K, equimolar composition and $p = 0.1013$ MPa. Comparison of experimental results (Exp.) with DISQUAC (DQ) calculations using the interaction parameters from Table 1

| System | $C_{pm}^{E}$/J mol$^{-1}$ K$^{-1}$ | | Ref. |
|---|---|---|---|
| | Exp. | DQ. | |
| $C_6H_5Cl + n\text{-}C_6$ | −0.80 | −0.82 | 59 |
| $C_6H_5Cl + n\text{-}C_7$ | −0.75 | −0.87 | 58 |
| $C_6H_5Cl + n\text{-}C_8$ | −1.19 | −1.11 | 59 |
| $C_6H_5Cl + n\text{-}C_{10}$ | −1.56 | −1.63 | 59 |
| $C_6H_5Cl + n\text{-}C_{12}$ | −2.26 | −2.29 | 59 |
| $C_6H_5Cl + n\text{-}C_{14}$ | −2.94 | −2.92 | 59 |
| $C_6H_5Cl + n\text{-}C_{16}$ | −3.52 | −3.68 | 59 |
| 1,2,4-trichlorobenzene + $n\text{-}C_6$ | −1.24 | −1.23 | 78 |
| 1,2,4-trichlorobenzene + $n\text{-}C_7$ | −0.91 | −1.11 | 63 |
| 1,2,4-trichlorobenzene + $n\text{-}C_9$ | −0.76 | −1.11 | 78 |
| 1,2,4-trichlorobenzene + $n\text{-}C_{10}$ | −0.66 | −0.92 | 78 |
| 1,2,4-trichlorobenzene + $n\text{-}C_{12}$ | −0.69 | −0.50 | 63 |
| 1,2,4-trichlorobenzene + $n\text{-}C_{16}$ | −1.89 | −2.05 | 63 |
| 1-chloronaphthalene + $n\text{-}C_6$ | 0.03 | 0.07 | 14 |
| 1-chloronaphthalene + $n\text{-}C_7$ | 0.82 | 0.95 | 14 |
| 1-chloronaphthalene + $n\text{-}C_8$ | 1.36 | 1.28 | 14 |
| 1-chloronaphthalene + $n\text{-}C_{12}$ | 1.68 | 1.69 | 14 |
| 1-chloronaphthalene + $n\text{-}C_{14}$ | 1.34 | 1.32 | 63 |

**TABLE 7**

DISQUAC (DQ) values on $S_{cc}(0)$, calculated using the interaction parameters from Table 1, and experimental (Exp.) results for aromatic halogenated compound(1) or 1-methylnaphthalene (1), or 1,2,4-trimethylbenzene (1) + alkane (2) mixtures at temperature $T$ and equimolar composition. The combinatorial contribution to $S_{cc}(0)^{-1}$, $S_{CC}(0)^{-1}(\text{comb})$, from the DISQUAC model is also included.

| System | $T$/K | $S_{CC}(0)$ Exp. | $S_{CC}(0)$ DQ | $S_{CC}(0)^{-1}(\text{comb})$ | Ref. |
|---|---|---|---|---|---|
| $C_6H_5Cl$ + heptane | 298.15 | | 0.380 | 0.096 | |
| | 323.15 | 0.359 | 0.359 | 0.096 | 67 |
| $C_6H_5Cl$ + decane | 323.15 | 0.267 | 0.318 | 0.385 | 80 |
| $C_6H_5Cl$ + tetradecane | 298.15 | | 0.307 | 0.797 | |
| $C_6H_5Br$ + heptane | 298.15 | | 0.394 | 0.069 | |
| $C_6H_5Br$ + tetradecane | 298.15 | | 0.344 | 0.730 | |
| $C_6H_5Br$ + cyclohexane | 298.15 | 0.371 | 0.371 | 0 | |
| 1-chloronapthalene + heptane | 298.15 | | 0.436 | 0.006 | |
| 1-chloronaphthalene + tetradecane | 298.15 | | 0.400 | 0.311 | |
| 1-chloronaphthalene + hexadecane | 298.15 | 0.371 | 0.385 | 0.453 | 62 |
| 1,2,4-trichlorobenzene + hexane | 293.15 | 0.565 | 0.555 | 0.011 | 110 |
| 1,2,4-trichlorobenzene + heptane | 298.15 | | 0.587 | 0.001 | |
| 1,2,4-trichlorobenzene + tetradecane | 298.15 | | 0.384 | 0.435 | |
| 1-methylnaphthalene + heptane | 298.15 | | 0.297 | 0.010 | |
| 1-methylnaphthalene + tetradecane | 298.15 | | 0.249 | 0.287 | |
| 1,2,4-trimethylbenzene + heptane | 298.15 | | 0.278 | 0.002 | |
| 1,2,4-trimethylbenzene + tetradecane | 298.15 | | 0.258 | 0.347 | |

**TABLE 8**

Excess molar functions: enthalpy, $H_m^E$, volume, $V_m^E$, and internal energy at constant volume $U_{Vm}^E$, at equimolar composition, 298.15 K and $p = 0.1013$ MPa for aromatic halogenated compound (1) + alkane (2) mixtures

| | C$_6$H$_5$Cl | | | C$_6$H$_5$Br | | | 1,2,4-trichlorobenzene | | | 1-chloronaphthalene | | |
|---|---|---|---|---|---|---|---|---|---|---|---|---|
| | $H_m^E$ / | $V_m^E$ / | $U_{Vm}^E$ / | $H_m^E$ / | $V_m^E$ / | $U_{Vm}^E$ / | $H_m^E$ / | $V_m^E$ / | $U_{Vm}^E$ / | $H_m^E$ / | $V_m^E$ / | $U_{Vm}^E$ / |
| | J mol$^{-1}$ | cm$^3$ mol$^{-1}$ | J mol$^{-1}$ | J mol$^{-1}$ | cm$^3$ mol$^{-1}$ | J mol$^{-1}$ | J mol$^{-1}$ | cm$^3$ mol$^{-1}$ | J mol$^{-1}$ | J mol$^{-1}$ | cm$^3$ mol$^{-1}$ | J mol$^{-1}$ |
| $n$-C$_6$ | 662[a] | −0.521[b] | 802 | 722[c] | −0.634[d] | 945 | 729[e] | −0.941[f] | 991 | 665[g] | −1.560[g] | 1092 |
| $n$-C$_7$ | 672[h] | −0.243[b] | 742 | 795[c] | −0.361[d] | 901 | 711[e] | −0.629[i] | 900 | 656[g] | −1.263[g] | 1031 |
| $n$-C$_8$ | | | | | | | | | | 658[g] | −1.052[g] | 979 |
| $n$-C$_9$ | 712[h] | 0.076[b] | 689 | 805[c] | −0.036[d] | 817 | 680[e] | −0.327[f] | 783 | | | |
| $n$-C$_{10}$ | | | | | | | 681[c] | −0.178[d] | 738 | 654[g] | | 899 |
| $n$-C$_{12}$ | | | | | | | | | | 630[g] | −0.616[g] | 828 |
| $n$-C$_{14}$ | 755[h] | 0.356[j] | 640 | 824[c] | 0.320[d] | 719 | 608[e] | 0.103[f] | 573 | | | |
| $n$-C$_{16}$ | | | | | | | | | | 580[g] | −0.397[g] | 712 |
| C$_6$H$_{12}$ | 699[k] | 0.351[l] | 577 | 740[c] | 0.281[m] | 640 | | | | | | |

[a][77]; [b][65]; [c][12]; [d][66]; [e][13]; [f][78]; [g][14]; [h][11]; [i][63]; [j][18]; [k][113]; [l][58]; [m][72]

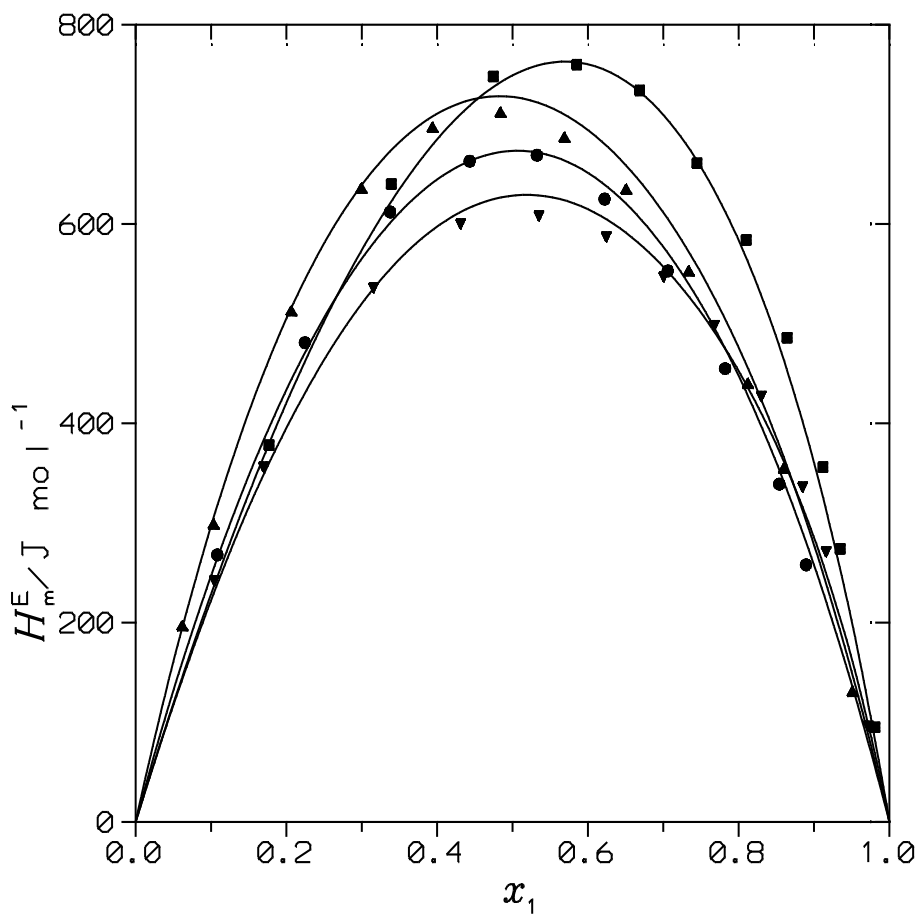

**Figure 1.** $H_m^E$ of aromatic halogenated compound (1) + *n*-alkane (2) mixtures at 298.15 K. Points, experimental results: (●), $C_6H_5Cl$ + heptane; (■), $C_6H_5Cl$ + tetradecane [11]; (▲), 1,2,4-trichlorobenzene + heptane; (▼), 1,2,4-trichlorobenzene + tetradecane [13]. Solid lines, DISQUAC calculations.

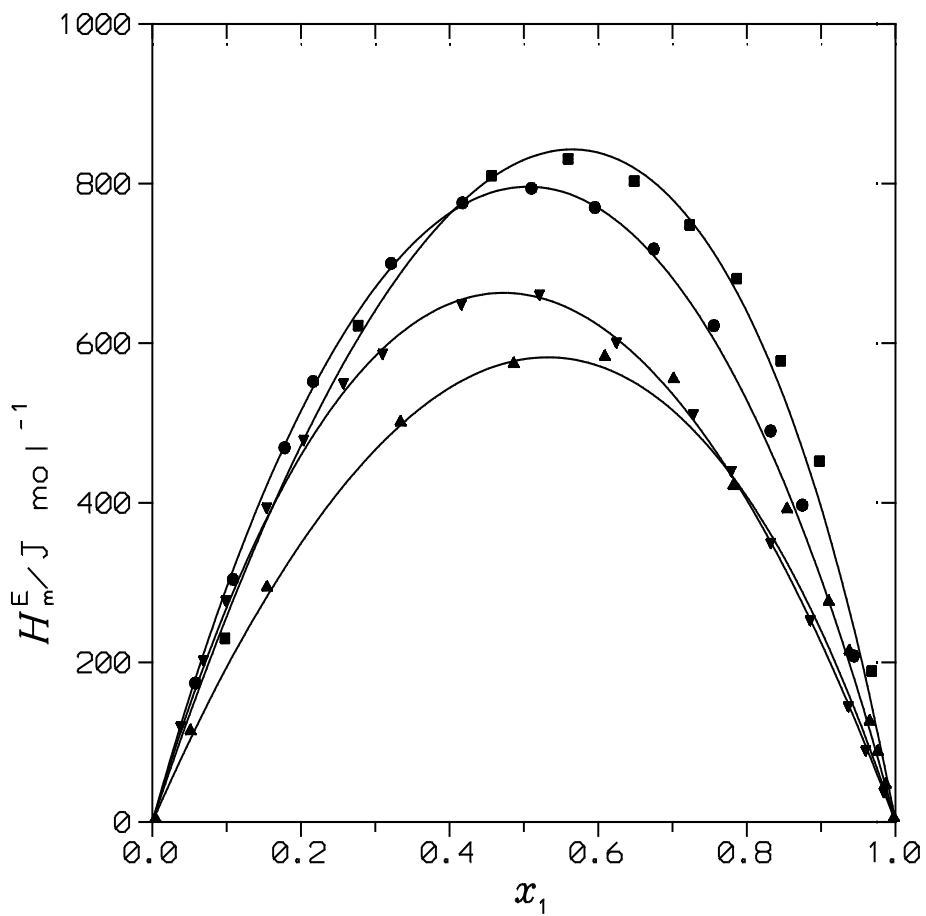

**Figure 2.** $H_m^E$ of aromatic halogenated compound (1) + *n*-alkane (2) mixtures at 298.15 K. Points, experimental results: (●), $C_6H_5Br$ + heptane; (■); $C_6H_5Br$ + tetradecane [12]; (▼), 1-chloronaphthalene + heptane; (▲), 1-chloronaphthalene + hexadecane [14]. Solid lines, DISQUAC calculations.

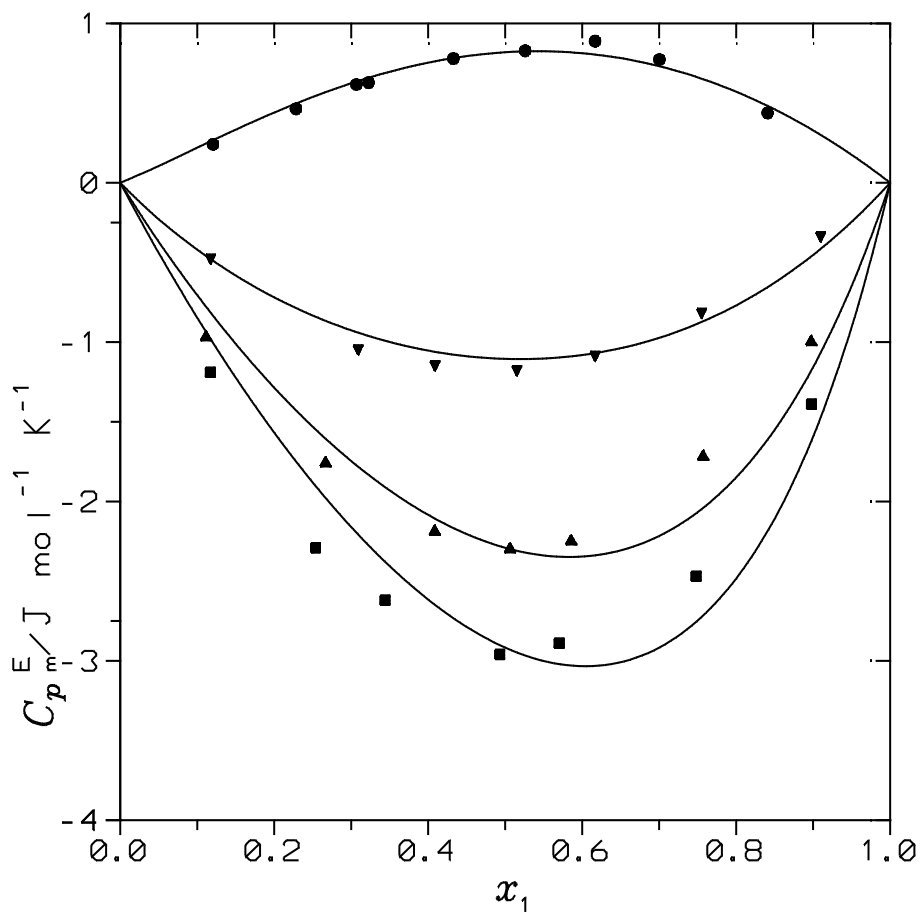

**Figure 3.** $C_{pm}^E$ of aromatic halogenated compound (1) + *n*-alkane (2) mixtures at 298.15 K. Points, experimental results: (▼), $C_6H_5Cl$ + octane; (▲), $C_6H_5Cl$ + dodecane; (■), $C_6H_5Cl$ + tetradecane [59]; (●), 1-chloronaphthalene + heptane [14]. Solid lines, DISQUAC calculations.

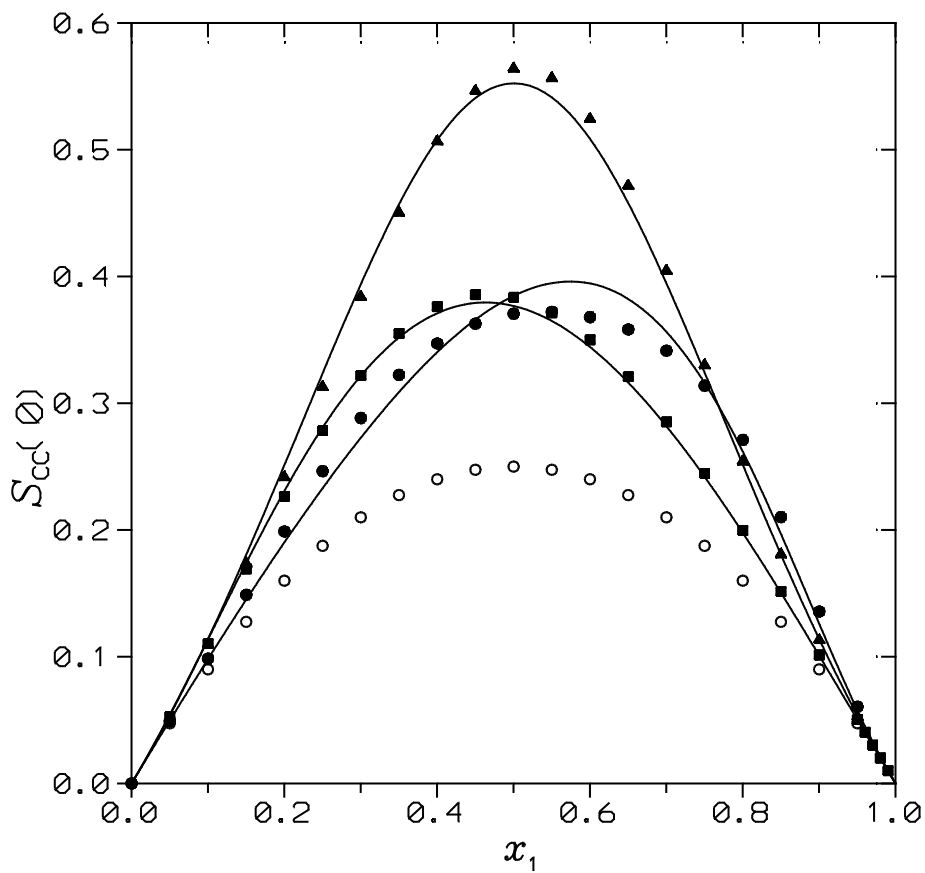

**Figure 4.** $S_{CC}(0)$ of aromatic halogenated compound (1) + alkane (2) mixtures. Full points, experimental results: (●), 1-chloronaphthalene + hexadecane ($T$ = 298.15 K) [62]; (■), $C_6H_5Br$ + cyclohexane ($T$ = 298.15 K) [110]; (▲), 1,2,4-trichlorobenzene + hexane ($T$ = 293.15 K) [110]. Open symbols, ideal mixture. Solid lines, DISQUAC calculations

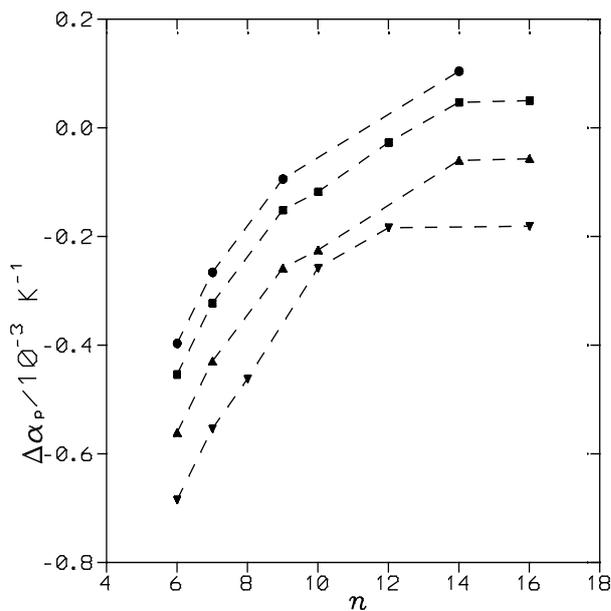

**Figure 5a.** Differences between thermal expansion coefficients, $\Delta\alpha_p = (\alpha_{p1} - \alpha_{p2})$ of mixture compounds in aromatic halogenated compound (1) + $n$-alkane (2) systems at 298.15 K vs. $n$, the number of C atoms of the $n$-alkane: (●), $C_6H_5Cl$; (■), $C_6H_5Br$; (▼), 1-chloronaphthalene; (▲), 1,2,4-trichlorobenzene (for references, see Table 2 and [49]. Dashed lines are for the aid of the eye.

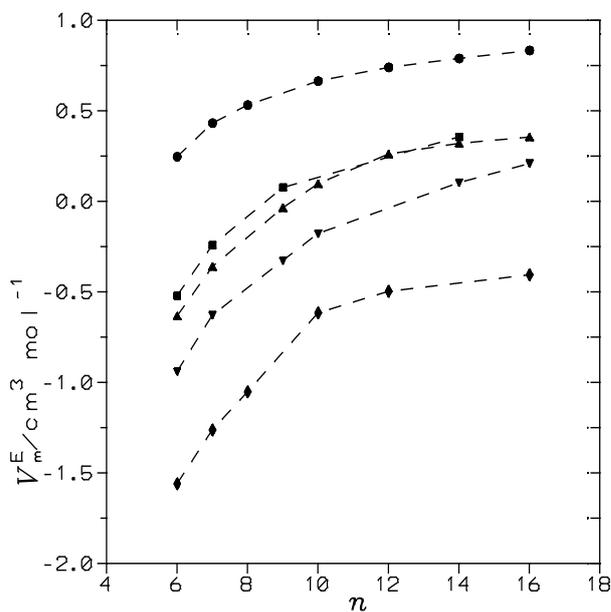

**Figure 5b** $V_m^E$ values at equimolar composition and 298.15 K of aromatic halogenated compound (1) + $n$-alkane (2) systems at 298.15 K vs. $n$, the number of C atoms of the $n$-alkane: (●), $C_6H_5F$ [60]; (■), $C_6H_5Cl$; (▲), $C_6H_5Br$; (♦), 1-chloronaphthalene; (▼), 1,2,4-trichlorobenzene (for references, see Table 8). Dashed lines are for the aid of the eye.

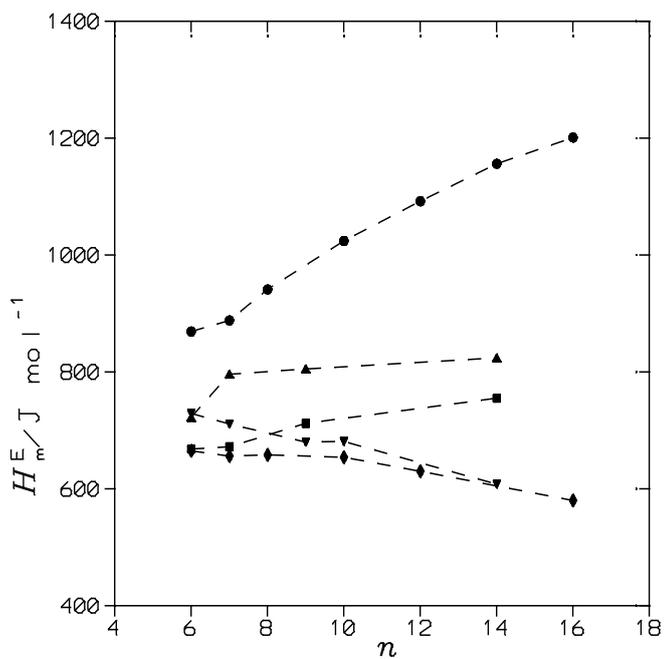 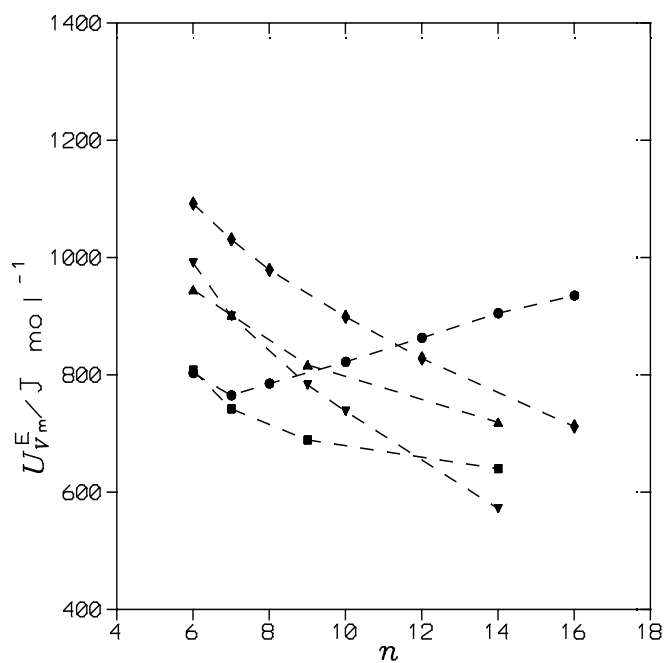

**Figure 6a**　　　　　　　　　　　　　　　　　　　　**Figure 6b**

**Figure 6.** $H_m^E$ (Figure 6a, for references, see Table 8, [60]) or $U_{Vm}^E$ (Figure 6b, for references, see Table 8 and [2]) of aromatic halogenated compound (1) + *n*-alkane (2) mixtures at equimolar composition and 298.15 K vs. *n*, the number of C atoms of the *n*-alkane: (●), $C_6H_5F$; (■), $C_6H_5Cl$; (▲), $C_6H_5Br$; (♦), 1-chloronaphthalene; (▼), 1,2,4-trichlorobenzene. Dashed lines are for the aid of the eye.

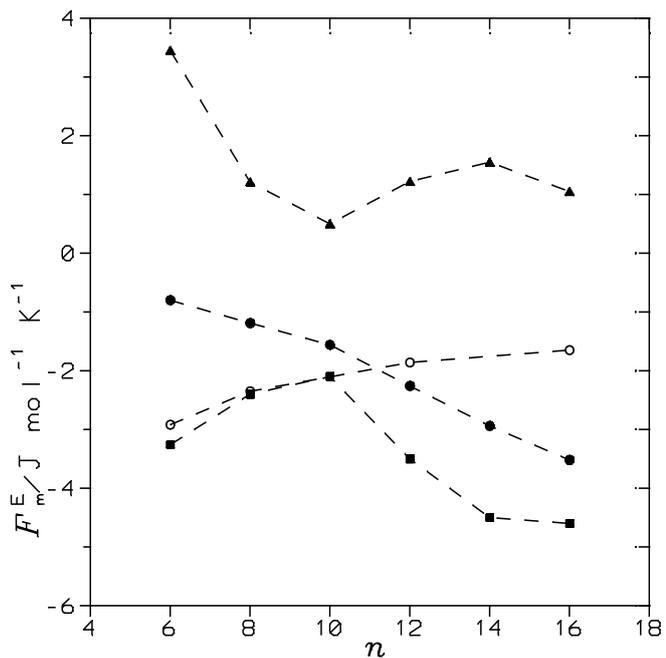 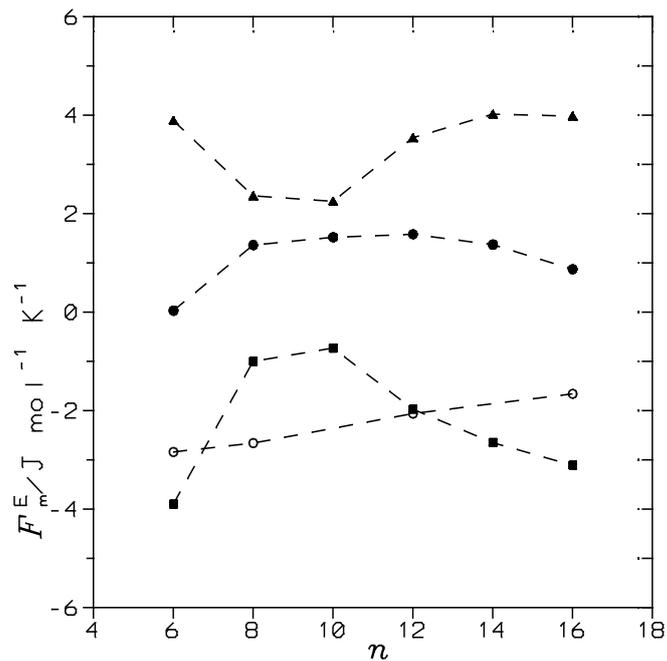

**Figure 7a**  **Figure 7b**

**Figure 7.** Molar excess heat capacities at equimolar composition and 298.15 K of $C_6H_5Cl$ (1) (Figure 7a) or of 1-chloronaphthalene (Figure 7b) (1) + alkane (2) mixtures vs. $n$, the number of C atoms of the alkane. Systems with $n$-alkanes: (●), $F_m^E = C_{pm}^E$ (for references, see Table 6]; (■), $F_m^E = C_{Vm}^E$. Open symbols, $F_m^E = C_{Vm}^E$ [16,18] for solutions including branched alkanes; (▲), represents the eos contribution to $F_m^E = C_{pm}^E$ (see text, eq. 15). Dashed lines are for the aid of the eye.

SUPPLEMENTARY MATERIAL

THERMODYNAMICS OF CHLOROBENZENE, OR BROMOBENZENE, OR 1-CHLORONAPHTHALENE OR 1,2,4-TRICHLOROBENZENE + ALKANE MIXTURES

Juan Antonio González,* Luis Felipe Sanz, Fernando Hevia, Isaías. García de la Fuente, and José Carlos Cobos

G.E.T.E.F., Departamento de Física Aplicada, Facultad de Ciencias, Universidad de Valladolid, Paseo de Belén, 7, 47011 Valladolid, Spain.

*corresponding author, e-mail: jagl@termo.uva.es; Fax: +34-983-423136; Tel: +34-983-423757

**TABLE S1**

Dispersive (DIS) and quasichemical (QUAC) interchange coefficients, $C_{ab,l}^{DIS}$ and $C_{ab,l}^{QUAC}$, for (a,b) contacts[a] in 1-methylnaphthalene or 1,2,4-trimethylbenzene + $n$-alkane mixtures ($l = 1$, Gibbs energy; $l = 2$, enthalpy; $l = 3$, heat capacity). The interaction energy parameter, $X_{12}$, in the Flory model is also listed.

| Alkane | DISQUAC interchange coefficients | | | | | | Flory parameter |
|---|---|---|---|---|---|---|---|
| | $C_{ab,1}^{DIS}$ | $C_{ab,2}^{DIS}$ | $C_{ab,3}^{DIS}$ | $C_{ab,1}^{QUAC}$ | $C_{ab,2}^{QUAC}$ | $C_{ab,3}^{QUAC}$ | $X_{12}$/J cm$^{-3}$ |
| 1-methylnaphthalene | | | | | | | |
| $n$-C$_6$ | 0.143 | 0.58[c] | 0.1 | 0 | 0 | 0 | |
| $n$-C$_7$ | 0.143 | 0.55 | 0.1 | 0 | 0 | 0 | 27.68 |
| $n$-C$_8$ | 0.143 | 0.51[c] | 0.1 | 0 | 0 | 0 | |
| $n$-C$_{10}$ | 0.143 | 0.433 | 0.1 | 0 | 0 | 0 | 22.28 |
| $n$-C$_{12}$ | 0.143 | 0.38 | 0.1 | 0 | 0 | 0 | 19.28 |
| $n$-C$_{14}$ | 0.143 | 0.343 | 0.1 | 0 | 0 | 0 | 17.51 |
| $n$-C$_{16}$ | 0.143 | 0.32 | 0.1 | 0 | 0 | 0 | 16.31 |
| 1,2,4-trimethylbenzene | | | | | | | |
| $n$-C$_6$ | 0.38 | 0.89[c] | 0.20[c] | 0 | 0 | 0 | |
| $n$-C$_7$ | 0.38 | 0.82 | 0.30 | 0 | 0 | 0 | 10.21 |
| $n$-C$_9$ | 0.38 | 0.70[c] | 0.40[c] | 0 | 0 | 0 | |
| $n$-C$_{10}$ | 0.38 | 0.65 | 0.50 | 0 | 0 | 0 | 7.92 |
| $n$-C$_{12}$ | 0.38 | 0.59 | 0.20[c] | 0 | 0 | 0 | 7.41 |
| $n$-C$_{14}$ | 0.38 | 0.59 | −0.06 | 0 | 0 | 0 | 7.38 |
| $n$-C$_{16}$ | 0.38 | 0.59 | −0.06[c] | 0 | 0 | 0 | 7.44 |

[a]type s = a, aliphatic, s = b, aromatic; [c]estimated value

**TABLE S2**

Molar excess enthalpies, $H_m^E$, at equimolar composition, 298.15 K and $p = 0.1013$ MPa for 1-methylnaphthalene (1) or 1,2,4-trimethylbenzene (1) + $n$-alkane (2) mixtures. Comparison of experimental results with values given by DISQUAC (DQ) or Flory models with interaction parameters listed on Table S1. $N$ is the number of data points and $dev(H_m^E)$ the relative deviation for $H_m^E$ (equation 11).

| alkane | N | $H_m^E$/J·mol$^{-1}$ | | $dev(H_m^E)$ | | | Ref. |
|---|---|---|---|---|---|---|---|
| | | Exp. | DQ | Exp. | DQ | Flory | |
| 1-methylnaphthalane | | | | | | | |
| $n$-C$_7$ | 15 | 759 | 772 | 0.002 | 0.037 | 0.029 | 17 |
| $n$-C$_{10}$ | 15 | 702 | 698 | 0.002 | 0.040 | 0.015 | 17 |
| $n$-C$_{12}$ | 15 | 649 | 652 | 0.004 | 0.034 | 0.025 | 17 |
| $n$-C$_{14}$ | 15 | 622 | 618 | 0.006 | 0.048 | 0.027 | 17 |
| $n$-C$_{16}$ | 29 | 597 | 600 | 0.008 | 0.064 | 0.070 | 17 |
| 1,2,4-trimethylbenzene | | | | | | | |
| $n$-C$_7$ | 16 | 275 | 276 | 0.020 | 0.029 | 0.023 | 17 |
| $n$-C$_{10}$ | 18 | 252 | 251 | 0.030 | 0.011 | 0.014 | 17 |
| $n$-C$_{12}$ | 16 | 250 | 243 | 0.034 | 0.016 | 0.026 | 17 |
| $n$-C$_{14}$ | 22 | 258 | 256 | 0.005 | 0.014 | 0.037 | 17 |
| $n$-C$_{16}$ | 18 | 268 | 266 | 0.029 | 0.016 | 0.052 | 17 |

**TABLE S3**

Isobaric molar excess heat capacities, $C_{pm}^{E}$, of 1-methylnaphthalene (1) or 1,2,4-trimethylbenzene (1) + $n$-alkane (2) mixtures at 298.15 K, equimolar composition and $p = 0.1013$ MPa. Comparison of experimental results (Exp.) with DISQUAC (DQ) calculations using the interaction parameters from Table S1

| System | $C_{pm}^{E}$/J mol$^{-1}$ K$^{-1}$ | | Ref. |
|---|---|---|---|
| | Exp. | DQ. | |
| 1-methylnaphthalene + $n$-C$_{12}$ | 0.52 | 0.58 | 16 |
| 1,2,4-trimethylbenzene + $n$-C$_7$ | 0.32 | 0.34 | 17 |
| 1,2,4-trimethylbenzene + $n$-C$_{10}$ | 0.66 | 0.65 | 17 |
| 1,2,4-trimethylbenzene + $n$-C$_{14}$ | −0.094 | −0.087 | 17 |

**TABLE S4**

Excess molar volumes, $V_m^E$, at 298.15 K and equimolar composition for aromatic halogenated compound (1), or 1,2,4-trimethylbenzene (1) + n-alkane (2) mixtures. Comparison of experimental (exp) results with values obtained from the Flory model with interaction parameters listed in Table 1.

| alkane | $V_m^E$/cm³mol⁻¹ | | Ref. |
|---|---|---|---|
| | Exp. | Flory | |
| $C_6H_5Cl$ + n-alkane | | | |
| n-C$_6$ | −0.521 | −0.263 | 65 |
| n-C$_7$ | −0.243 | 0.060 | 65 |
| n-C$_9$ | 0.076 | 0.378 | 65 |
| n-C$_{14}$ | 0.356 | 0.669 | 18 |
| cyclohexane | 0.351 | 0.347 | 58 |
| $C_6H_5Br$ | | | |
| n-C$_6$ | −0.634 | −0.529 | 66 |
| n-C$_7$ | −0.361 | −0.119 | 66 |
| n-C$_9$ | −0.036 | 0.229 | 66 |
| n-C$_{14}$ | 0.320 | 0.632 | 66 |
| cyclohexane | 0.281 | 0.196 | 72 |
| 1-chloronaphtahlene + n-alkane | | | |
| n-C$_6$ | −1.560 | −1.636 | 14 |
| n-C$_7$ | −1.263 | −1.148 | 14 |
| n-C$_{12}$ | −0.616 | −0.409 | 14 |
| n-C$_{16}$ | −0.397 | −0.210 | 14 |
| | −0.406 | | 63 |
| 1,2,4-trichlorobenzene + n-alkane | | | |
| n-C$_6$ | −0.941 | −1.178 | 78 |
| n-C$_7$ | −0.629 | −0.737 | 63 |
| n-C$_9$ | −0.327 | −0.317 | 78 |
| n-C$_{10}$ | −0.178 | −0.219 | 78 |
| n-C$_{14}$ | 0.103 | 0.181 | 78 |
| 1,2,4-trimethylbenzene + n-alkane | | | |
| n-C$_7$ | −0.207 | −0.313 | 17 |
| n-C$_{10}$ | 0.0596 | 0.118 | 17 |
| n-C$_{14}$ | 0.145 | 0.218 | 17 |

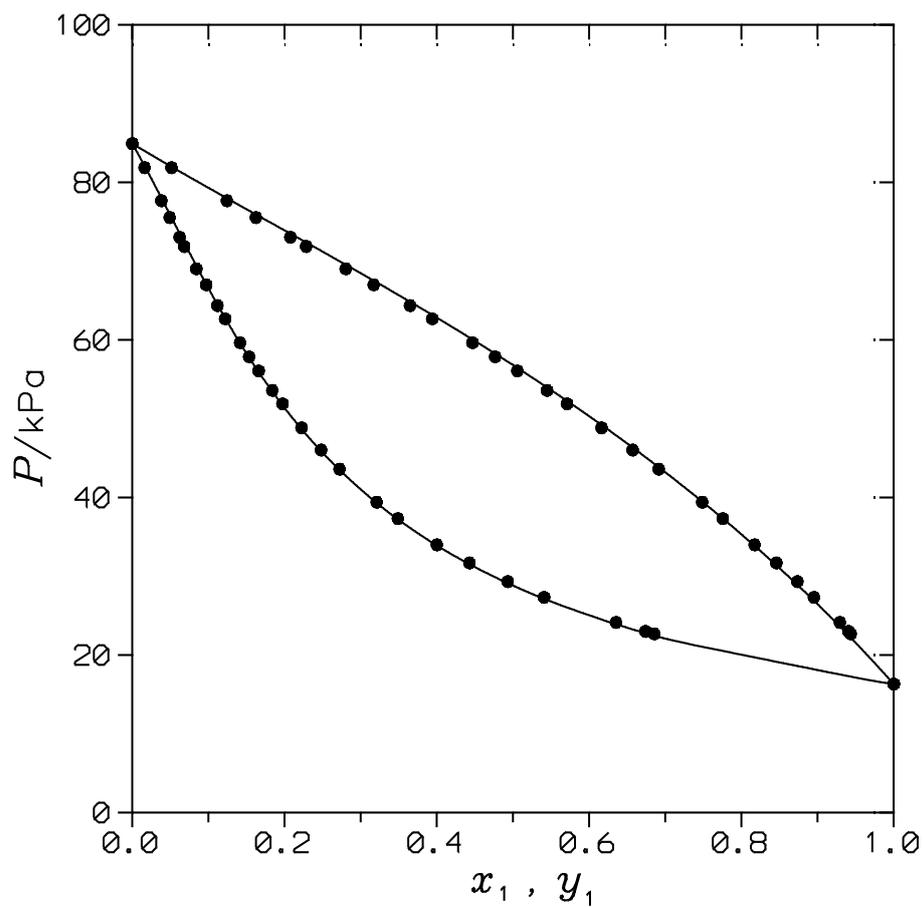

**FIG. S1**  VLE for the chlorobenzene (1) + cyclohexane (2) mixture at 348.15 K. Points, experimental results [108]. Solid lines, DISQUAC calculations using interaction parameters from Table 1.

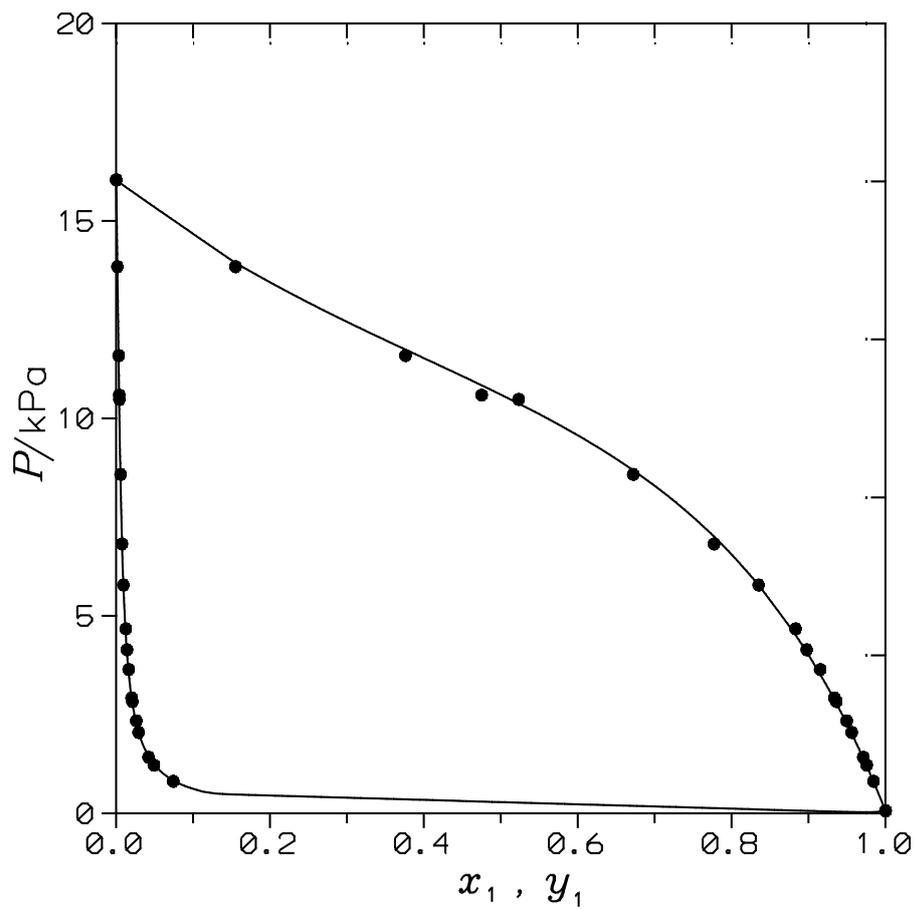

**FIG. S2** VLE for the 1,2,4-trichlorobenzene (1) + hexane (2) mixture at 293.15 K. Points, experimental results [110]. Solid lines, DISQUAC calculations using interaction parameters from Table 1.

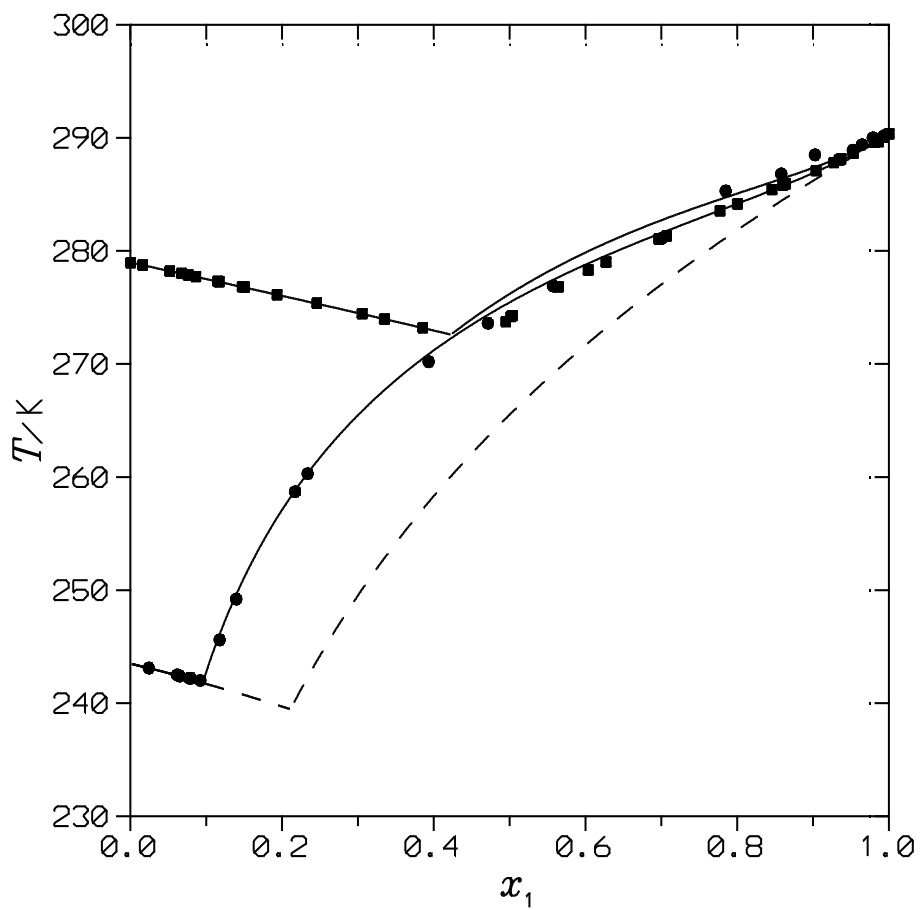

**FIG. S3** SLE for the 1,2,4-trichlorobenzene (1) + decane (2) (●) or + tetradecane (2) (■) mixtures Points, experimental results [53]. Solid lines, DISQUAC calculations using interaction parameters from Table 1. Dashed lines, IDEAL solubility model for the system with decane.